\documentclass[lettersize,journal]{IEEEtran}
\usepackage{amsmath,amsfonts}
\usepackage{array}
\usepackage[caption=false,font=normalsize,labelfont=sf,textfont=sf]{subfig}
\usepackage{textcomp}
\usepackage{stfloats}
\usepackage{url}
\usepackage{verbatim}
\usepackage{graphicx}
\usepackage{cite}
\usepackage[utf8]{inputenc}
\usepackage{url}
\usepackage{amsfonts}
\usepackage{nicefrac}
\usepackage{xcolor}
\usepackage{color}
\usepackage{picinpar}
\usepackage{cutwin}
\usepackage{url}
\usepackage{booktabs}
\usepackage{multirow}
\usepackage{times}
\usepackage{epsfig}
\usepackage{bbm}
\usepackage{enumerate}
\usepackage{algorithm}
\usepackage[noend]{algpseudocode}
\usepackage{bm}
\usepackage{amsthm}
\usepackage{mathtools}
\usepackage{dsfont}
\usepackage{bbding}
\usepackage{enumitem}
\usepackage[hang,flushmargin]{footmisc}
\usepackage{wrapfig}
\usepackage{hyperref}

\newcommand\shortsection[1]{{\noindent\bf #1.}}
\newcommand{\trainset}{\mathcal{D}_\text{tr}}
\newcommand{\poisonsubset}{\mathcal{D}_\text{p}}
\newcommand{\cleansubset}{\mathcal{D}_\text{c}}
\newcommand{\poisonset}{\mathcal{D}'_\text{p}}
\DeclareMathOperator*{\argmin}{argmin}

\begin{document}

\title{Invisibility Cloak: Disappearance under Human Pose Estimation via Backdoor Attacks}

\author{Minxing Zhang, Wenshu Fan,~\IEEEmembership{Student Member,~IEEE,} Wenbo Jiang,~\IEEEmembership{Member,~IEEE,} \\ Shui Yu,~\IEEEmembership{Fellow,~IEEE,} Michael Backes,~\IEEEmembership{Fellow,~IEEE,} Xiao Zhang$^*$

\thanks{M. Zhang, M. Backes and X. Zhang are with CISPA Helmholtz Center for Information Security, 66123 Saarbrücken, Germany (e-mail: minxing.zhang@cispa.de; director@cispa.de; xiao.zhang@cispa.de).}
\thanks{W. Fan and W. Jiang are with the School of Computer Science and Engineering, University of Electronic Science and Technology of China, Chengdu 610056, China (e-mail: fws@std.uestc.edu.cn; wenbo\_jiang@uestc.edu.cn).}
\thanks{S. Yu is a Professor of the School of Computer Science in the Faculty of Engineering and Information Technology at University of Technology Sydney, Sydney, Australia (e-mail: Shui.Yu@uts.edu.au).}
\thanks{$^*$Xiao Zhang is the corresponding author.}
}



\maketitle

\begin{abstract}
Despite being significant in autonomous systems, Human Pose Estimation (HPE)'s potential risks to adversarial attacks have not received comparable attention with image classification or segmentation tasks.
In this paper, we study the vulnerability of HPE systems to disappearance attacks, where the attacker aims to subtly alter the HPE training process via backdoor techniques so that any input image with some specific trigger will not be recognized as involving any human pose.
As humans are typically at the center of HPE systems, a successful attack will severely threaten pedestrians' lives if a self-driving car incorrectly understands the front scene.
To achieve the adversarial goal of disappearance, we propose \emph{IntC}, a general framework to craft an invisibility cloak in the HPE domain.
By designing target HPE labels that do not represent any human pose, we propose three specific backdoor attacks based on our IntC framework.
IntC-S and IntC-E, respectively designed for regression- and heatmap-based HPE techniques, concentrate the keypoints of triggered images in a tiny, imperceptible region.
Further, to improve the attack's stealthiness, IntC-L designs the target poisons to capture the label outputs of typical landscape images without a human involved, achieving disappearance and reducing detectability simultaneously.
Extensive experiments demonstrate the effectiveness and generalizability of our IntC methods in achieving the disappearance goal.
By revealing the vulnerability of HPE to disappearance and backdoor attacks, we hope our work can raise awareness of the potential risks when HPE models are deployed in real-world applications.
\end{abstract}


\section{Introduction}
\label{section: introduction}

Due to the success of deep neural networks (DNNs), human pose estimation (HPE), an interesting topic in computer vision, has evolved significantly in recent years.
Specifically, HPE aims to predict the locations of a person's keypoints (e.g., body joints), referred to as the HPE label, that determine the human pose in a given human-involved image.
Existing approaches for learning HPE models can be mainly divided into two categories, i.e., regression-based and heatmap-based.
Regression-based technique, which aims to predict the position of each keypoints directly, starts the DNN era of HPE~\cite{DeepPose}.
Later studies consider the HPE task as a sequence of heatmap estimation problems.
Instead of directly predicting the keypoint positions, the heatmap-based method predicts a heatmap for each keypoint, where each pixel in a heatmap represents the possibility of the corresponding keypoint being located at this position.
Such a focus shift to heatmaps is motivated by the observations that more spatial location information can be preserved in heatmaps, smoothing the training process.
Human pose estimation is significant for automatically controlled applications such as self-driving and robotics~\cite{KBA19,XYWO20,CYLDZZ21,LRY22,PMP23}.
For instance, HPE can be integrated into a self-driving car's decision-making by processing the captured front scene to predict the human pose.
Intuitively, the estimated poses are available for identifying front persons, e.g., pedestrians.
Besides, HPE is expected to provide more functionalities, e.g., the predicted specific poses are used to understand the intention and predict the critical pose changes of a person~\cite{GWH19,KRCRWMJF21,ZSGMSQLCCZLA21}.
While enjoying the benefits of HPE models, paying attention to their potential security risks is essential, especially given that HPE applications are typically human-centered.

However, only a few existing works explore HPE vulnerabilities, and they all focus on evasion attacks~\cite{JSKJ19,CVAZ22,LZWYM23}.
These works consider an adversary goal: mislead an HPE model to output incorrect poses using adversarial examples.
We argue that the severity of such a threat model is limited in HPE applications since the targeted incorrect predictions are still regarded as human-involved.
In this work, we propose to study disappearance attacks against HPE models.
Here, the notion of \emph{disappearance} does not refer to removing the input image pixels corresponding to a person but to causing a person in an input image to disappear with respect to the predicted HPE label.
Compared with the threat model studied in existing works, such an adversarial goal may lead to more severe consequences for human-centered HPE applications.
For instance, a self-driving car will be misled to keep forward by a disappearance attack despite a pedestrian being in front, which directly threatens the pedestrian's life.
Therefore, we hope our work can enhance the community's awareness of HPE vulnerabilities and potential risks when deploying HPE techniques in real-world systems.

\subsection{Our Work}
\label{subsection: our work}

We focus on backdoor attacks to accomplish disappearance, which can explicitly establish an adversarial connection between a specific trigger pattern and any desired HPE label.
As the first exploration of disappearance attacks in HPE, our work prefers such a straightforward method.
When a small portion of the training dataset is poisoned by a specific input trigger and carefully designed HPE labels, the induced backdoored model will produce non-human predictions on any triggered images.
The core of our work lies in designing target HPE labels to capture the disappearance goal properly.

\shortsection{Challenges}
Existing backdoor attacks mainly focus on classification tasks~\cite{CLLLS17,YLZZ19,LMBL20,NT20}.
Pursuing a more strict output requirement, disappearance attacks against HPE are more challenging than traditional backdoor attacks on classification models due to the distinctions in the structure of the output label space.
Specifically, a desired but incorrect human pose is a successful case for backdoor attacks.
However, it does not satisfy the disappearance requirement as the prediction still indicates a person.
On the contrary, our disappearance attack expects that a backdoored HPE model's predictions on triggered images do not represent any human pose.
Consequently, it is essential for our work to design proper non-person labels to connect with some specific trigger pattern using backdoor techniques.
Fortunately, the flexibility of HPE label space provides the feasibility for such a challenging task.
Concretely, each person has several keypoints, and each keypoint can be located at any place in an image, even if some positions are not common but still valid.
Therefore, numerous keypoint combinations are available, which also include non-human ones.
By proper label designs, we propose the first disappearance attack against HPE -- ``\textbf{In}visibili\textbf{t}y \textbf{C}loak'' (\emph{IntC}), which is also the first backdoor attack applied in the HPE domain.

\shortsection{Methodology}
The critical challenge is how to design appropriate poison labels that satisfy the disappearance requirement.
First, we conduct a naive attempt by directly setting the poison label as the label of a new clean image, similar to the traditional backdoor attacks.
However, the clean label still represents a human pose, which violates the expected disappearance goal.
Accordingly, we regard this method as the baseline, denoted as \emph{IntC-B}.
To achieve the goal of disappearance, we propose three specific backdoor label designs for HPE models. The first two designs concentrate all the keypoints in a tiny area so that they are hard to notice, while the third one expects triggered images to be regarded as landscape images, i.e., no person is involved.
In particular, we devise \emph{IntC-S} for regression-based methods, where the keypoints of a person are located at a single point, and develop \emph{IntC-E} that sets the heatmaps of keypoints as empty for heatmap-based methods.
To improve the attack stealthiness, we additionally propose \emph{IntC-L}, which captures the average information of landscape images as the poison label.
In summary, our novel label designs allow the IntC attacks to be applied in different scenarios to accomplish the disappearance task.

\shortsection{Evaluation}
We conduct comprehensive experiments to examine the performance of our IntC attacks.
Specifically, we consider a wide variety of pose estimation techniques, including DeepPose~\cite{DeepPose}, ChainedPredictions~\cite{ChainedPredictions}, HRNet~\cite{HRNet} and DEKR~\cite{DEKR}, and different benchmark datasets, such as COCO~\cite{MSCOCO}, MPII~\cite{MPII} and CrowdPose~\cite{CrowdPose}.
To quantitatively measure the attack performance, we involve two evaluation metrics, \emph{Utility} and \emph{Attack Success Rate} (ASR):
Utility measures the performance of an HPE model on the clean testing data, while ASR measures the attack performance on the triggered testing data.
Our results demonstrate the effectiveness of our IntC attacks in achieving high ASR while maintaining the model utility.
We also visualize the HPE label predictions, corroborating that the disappearance goal is indeed achieved on triggered test images.
In addition, we study the effectiveness of several possible backdoor defenses~\cite{CWW22,ZWZW23,GXWCRN19,DAR20,LDG18,LLKLLM21}, and discuss the potential adaptive defense directions.
Although existing defenses bring mitigation to some extent, our IntC attacks still achieve a high success rate, particularly if compared with the baseline method.
We hope our findings can increase the community's awareness of the potential security risks of HPE techniques when employed for real-world applications.

\shortsection{Contributions}
Our key contributions are outlined below:
\begin{itemize}
    \item We are the first to study the security risks of HPE models in the presence of disappearance attacks.
    \item Various backdoor attacks are proposed to realize the adversarial goal of disappearance by novel targeted label designs, with considerations of attack stealthiness.
    \item Through extensive experiments, our methods show effectiveness and generalizability in achieving our disappearance goal for HPE models.
\end{itemize}

\section{Background and Related Works}
\label{section: related works}

\begin{figure}[!t]
    \centering
    \subfloat{
        \centering
        \label{subfigure: HPE}
        \includegraphics[width=0.20\textwidth]{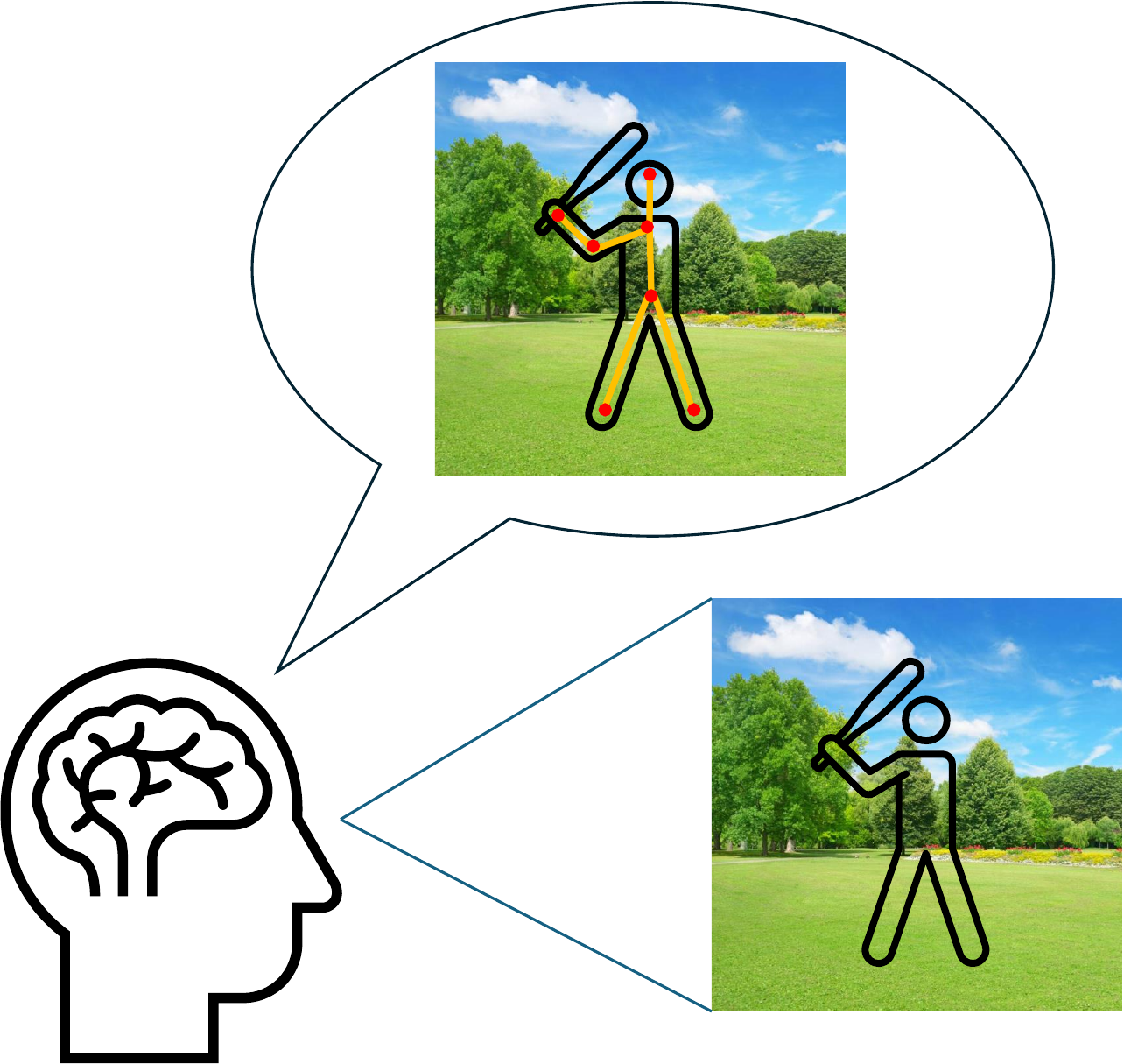}
    }
    \hspace{0.1in}
    \subfloat{
        \centering
        \label{subfigure: IntC}
        \includegraphics[width=0.20\textwidth]{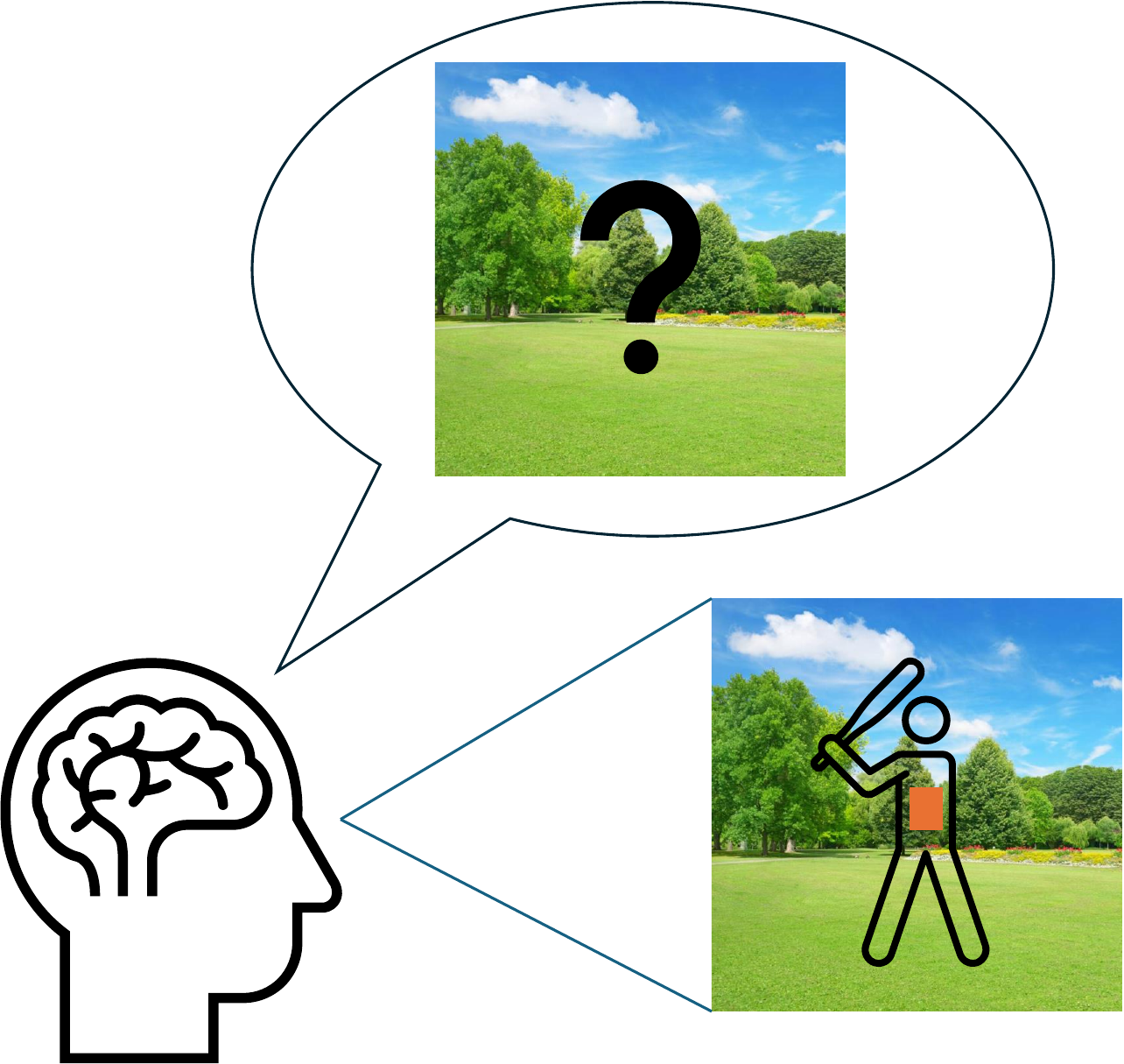}
    }
    % \vspace{-0.07in}
    \caption{Illustration of HPE (\textbf{left}) and our IntC attack (\textbf{right}).}
    % \vspace{-0.1in}
    \label{figure: HPE and our attack}
\end{figure}

\shortsection{Human Pose Estimation}
As depicted in Figure~\ref{subfigure: HPE}, HPE predicts the 2D positions of a single person's keypoints in an image, by which the human pose is determined.
To be more specific, the input of an HPE model is an image $x$ containing a person, and the output is the corresponding positions of the person's keypoints $y$, which can be formulated as:
\begin{align}
\label{equation: human pose estimation}
    \mathcal{M}_\text{HPE}(x) = y = \{p_1, \cdots, p_n\},
\end{align}
where $n$ is the number of relevant keypoints and $p_i$ is the position of the $i$-th keypoint.
In other words, each HPE label is a set of values indicating the positions of a person's keypoints.

Due to the advancement of deep learning, many works have proposed to employ DNNs to improve the accuracy of HPE~\cite{DeepPose,StackedHourglass,ChainedPredictions,HRNet}.
In taxonomy, these studies can be categorized as either regression-based or heatmap-based methods.
Specifically, regression-based methods directly predict the location of each keypoint.
Toshev et al. first propose to use DNNs to solve this regression problem, i.e., DeepPose~\cite{DeepPose}.
Inspired by DeepPose, some following works explore HPE as a regression task by taking advantage of the DNN development, e.g., adopting a re-parameterized and bone-based representation~\cite{SSLW17}, using soft-argmax function to convert feature maps into keypoint locations~\cite{LTP19}, and involving transformers~\cite{LWZXXT21}. 
Another line of work proposed to transfer HPE into a heatmap-based problem~\cite{ChainedPredictions,WRKS16,StackedHourglass,CYOMYW17,YLOLW17,HRNet,YXGYZSW21,YFHLZCW21,CXWSHZ20}, as heatmaps can preserve the spatial location information better and make the training process smoother~\cite{ZWCYZSKS}.
In particular, each pixel value in a heatmap indicates the probability that the corresponding keypoint lies in the position.
In addition, some recent works study the multi-person pose estimation~\cite{BUCTD,MIPNet,KAPAO,DEKR}, i.e., human poses of multiple people are estimated in a single image.
A primary study is conducted to explore the generalizability of our work in this more complex scenario in Section~\ref{subsection: multi-person pose estimation}.
Human poses in the 3D spatial location have also been explored in prior literature~\cite{ARS10,YWYRP11,MHRL17,TRA17}.
As this topic is less studied in the field and is regarded as the HPE technique followed by a matching module~\cite{CR17}, we decided not to include it and left the exploration in 3D HPE as future work.

\shortsection{Attacks Against Human Pose Estimation}
Unlike image classification, the attacks studied in the HPE domain are much fewer, whereas evasion attacks are the main focus.
For instance, Shah et al. conducted a comprehensive study of the HPE robustness against adversarial examples, by which they proposed insight into the design choices of pose-estimation systems to shape future work~\cite{JSKJ19}.
Chawla et al. showed how additive imperceptible perturbations could change predictions to increase the trajectory drift and catastrophically alter its geometry~\cite{CVAZ22}.
Liu et al. proposed local imperceptible adversarial examples on HPE networks, which reformulated imperceptible adversarial examples on keypoint regression into a constrained maximum allowable attack~\cite{LZWYM23}.
However, the potential HPE risks of these works are limited as the misestimated poses still represent some persons.
Therefore, this paper proposes the first disappearance attack against HPE, which explores HPE's vulnerability by causing the person in an image to disappear.

\shortsection{Disappearance Attacks}
The concepts of disappearance attacks and invisibility cloaks have been investigated in various computer-vision tasks~\cite{SEEFLRTPK18,YXLYRWWZ18,WLDG20,ZD23}, including object detection and facial recognition.
Existing works manage to achieve the disappearance goal by misclassification solutions.
For instance, a target model is misled to predict an incorrect category (i.e., identity) for the detected object (i.e., a human face image), where the predicted label is expected to be different from the ground truth.
However, a different predicted HPE label still suggests that a person is in the captured image, which does not support our disappearance goal.
This gap is derived from the fact that HPE is not a simple classification task, which motivates our work that aims to design proper HPE labels that do not represent any human pose.

\shortsection{Backdoor Attacks}
Compared to evasion attacks, backdoor attacks have the advantage of constructing an adversarial connection between a certain trigger and any desired label by poisoning strategies while minimizing the influence on the predictions of clean inputs.
The majority of existing backdoor attacks focused on classification tasks~\cite{YLZZ19,SSP20,WSRVASLP20,LMBL20,CT21,STKP21,JLG22,KLMGSIM23}, where an adversary can strategically attach triggers to a few training data and modify the corresponding labels.
When trained on the poisoned data, the target model can be misled to predict the expected class on any triggered image.
In contrast, our work generalizes the backdoor technique to the HPE domain, aiming to achieve the disappearance goal.
The disappearance feasibility is derived from the flexible HPE label space.
Specifically, each person has several keypoints, each of which can be located at any position in an image.
In that case, by probably designing poison labels, a backdoored HPE model can provide a prediction that indicates no person in a triggered image.
In other words, the core of the backdoor attack against HPE is how to design the desired labels.
By proposing novel label designs, our work brings a new attack surface to explore the security risks in the HPE domain.

\section{Threat Model}
\label{section: threat model}

\shortsection{Adversarial Goal}
Informally, the adversary aims to cause the person to ``disappear'' in the query image when predicted by the backdoored HPE model.
As the first work that explores the disappearance in the HPE domain, we will conceptually introduce the notion of disappearance in this section and provide the specific objectives of our backdoor attacks in Section~\ref{section: our attack}.
In particular, disappearance is carried out not by erasing the pixels of the corresponding person in the input image but by misleading the target HPE model to output the label prediction that does not represent any person.
By generalizing backdoor attacks to the HPE domain, we can explicitly connect triggers and the labels that indicate no person.
Benefiting from the flexible HPE label space, our work proposes novel label designs to implement the disappearance task, which will be detailed in Section~\ref{section: our attack}.

Making a person disappear at the prediction level of HPE models can induce severe risks in practice, as HPE is prominently adopted in autonomous systems.
For instance, HPE modules are deployed in self-driving cars to process the captured front scenes~\cite{BGCACFJBPMSS19}.
The HPE module could capture this information when a pedestrian is in front.
If (a part of) the estimated person's body is on the driving route, the car has to make the appropriate decision, e.g., stopping or bypassing; otherwise, the car will keep going forward.
Since the driving decision relies on the HPE estimation, disappearance attacks threaten pedestrians' safety.
In addition, further downstream tasks might enhance the HPE application, e.g., a module that follows to explain the exact meaning of the detected human pose.
However, these add-ups are out of the standard scope of HPE.
Thus, we leave the study as future work.

\shortsection{Adversarial Knowledge}
Our work aims to accomplish the disappearance goal using backdoor techniques.
Following existing backdoor works, we impose the following mild assumptions on the poisoning adversary, i.e., can access and manipulate a small portion of clean training dataset~\cite{SHNSSDG18,GLDG19}.
To be more specific, the attacker first replaces a few clean samples for training the HPE model with the corresponding trigger-attached images and properly designed HPE labels.
The model trained on such poisoned training dataset, denoted as \emph{backdoored HPE model} throughout this paper, will provide a non-human prediction on any input image with the same trigger pattern.
To maintain the reasonable level of attack stealthiness, the poisoning number and the attached trigger size are preferred to be as small as possible, which will be discussed in Section~\ref{subsection: hyperparameters in our attack}.
Besides the poisoning access, we do not make any extra assumptions or requirements.

Although poisoning access is a commonly used setting in backdoor attacks, it would be helpful to discuss the practical feasibility of such an attack ability.
The first potential possibility is intrusion techniques~\cite{LLLT13,KGVK19}.
Specifically, when an attacker evades or breaks the protection of the target model's training data, they can easily modify the training data according to their expectations.
Additionally, a large-scale dataset is normally collected from the internet~\cite{LAION400M,MSCOCO}.
This decision provides the possibility of poisoning training data as the attacker can post their designed data online.
Besides, cloud computing is a prevalent service where big companies provide individual users with computational resources, of which privacy and security concerns are worth heated discussions~\cite{WLYHKTF10,DWC10}.
For instance, a malicious service provider can easily access and modify the users' stored data.
Note that the above discussion serves to understand the access to poisoning training data.
Besides, the exact skills will not be discussed in the paper, as our work focuses on the disappearance task in HPE.
In the following, we will introduce how to achieve the disappearance goal with our novel label designs in Section~\ref{section: our attack}.

\section{Attack Methodology}
\label{section: our attack}

\subsection{Invisibility Cloak}
\label{subsection: invisibility cloak}

To achieve the disappearance goal in the HPE domain, we propose our attack ``\textbf{In}visibili\textbf{t}y \textbf{C}loak'' (\textbf{IntC}).

\shortsection{Challenge}
Our attack generalizes backdoor attacks to HPE models.
Directly transferring the existing backdoor attacks against classifiers to the HPE domain cannot achieve the disappearance goal.
This is because the existing works simply replace the original labels with other ones.
However, in the HPE task, other human poses still represent persons, i.e., some persons still appear in the prediction.
In other words, disappearance requires the backdoored HPE model to output non-person predictions on triggered images.
Fortunately, the HPE label space is flexible. There are several keypoints for a person, and each keypoint can be located anywhere, which makes implementing the disappearance task feasible.
Specifically, by designing a proper label that does not represent any human pose, the person will disappear in the prediction of a backdoored HPE model by using the trigger.
Therefore, the label design is the core of our work.
Regarding the trigger, we use the classic design, i.e., attaching a small patch to the poison images.
This decision is because if our attack performs well with such a straightforward trigger, our work could provide a new attack surface for HPE.
In addition, it is interesting to study the combination with other more advanced trigger techniques~\cite{LMBL20,LLWLHL21,KLMGSIM23} in future work.

\begin{algorithm}[!t]
\caption{Training Data Poisoning of Invisibility Cloak}
\label{algorithm: invisibility cloaks}
\begin{algorithmic}
\Function{InvisibilityCloak}{training dataset $\trainset$, poison rate $\alpha$, trigger injection function $\tau$, label transformation function $\varphi$}
    % \\
    \State $\cleansubset$, $\poisonsubset$ $\gets$ Split training data $\trainset$ by poison rate $\alpha$
    \State \Comment{$|\poisonsubset|=\alpha\cdot|\trainset|$ and $\poisonsubset=\mathcal{X}_\mathrm{p}\times\mathcal{Y}_\mathrm{p}$.}
    % \\
    \If{``IntC-B''}
        \State $\varphi(\cdot)$ $\gets$ $\text{IntC-B}(\cdot)$ \Comment{Algorithm~\ref{algorithm: IntC-B}.}
    \ElsIf{``IntC-S''}
        \State $\varphi(\cdot)$ $\gets$ $\text{IntC-S}(\cdot)$ \Comment{Algorithm~\ref{algorithm: IntC-S}.}
    \ElsIf{``IntC-E''}
        \State $\varphi(\cdot)$ $\gets$ $\text{IntC-E}(\cdot)$ \Comment{Algorithm~\ref{algorithm: IntC-E}.}
    \ElsIf{``IntC-L''}
        \State $\varphi(\cdot)$ $\gets$ $\text{IntC-L}(\cdot)$ \Comment{Algorithm~\ref{algorithm: IntC-L}.}
    \EndIf
    % \\
    \State $\mathcal{X}'_\mathrm{p}$ $\gets$ $\tau(\mathcal{X}_\mathrm{p})$ \Comment{Inject triggers by $\tau(\cdot)$.}
    
    \State $\mathcal{Y}'_\mathrm{p}$ $\gets$ $\varphi(\mathcal{Y}_\mathrm{p})$ \Comment{Transform labels by $\varphi(\cdot)$.}
    % \\    
    \State $\poisonset$ $\gets$ $\mathcal{X}'_\mathrm{p}\times\mathcal{Y}'_\mathrm{p}$
    
    \State \Return Poisoned training data $\cleansubset\cup\poisonset$
\EndFunction
\end{algorithmic}
\end{algorithm}

\shortsection{Problem Formulation}
In the following, we will formulate the problem that our work aims to solve.
As Figure~\ref{subfigure: IntC} shows, IntC misleads the target model to think there is no person in an image by using the trigger.
Specifically, given a clean training dataset $\trainset \subseteq \mathcal{X} \times \mathcal{Y}$, where $\mathcal{X}$ denotes the input image space, and $\mathcal{Y}$ represents the label space of ground-truth keypoints.
The attacker aims to poison a small part of $\trainset$ with a poison rate $\alpha \in (0,1)$. Let $\cleansubset\subseteq\trainset$ and $\poisonsubset\subseteq\trainset$ denote the set of clean examples that will be poisoned and the remaining unchanged set, respectively, where $|\poisonsubset| = \alpha \cdot |\trainset|$, $\cleansubset\cup\poisonsubset=\trainset$ and $\cleansubset\cap\poisonsubset=\emptyset$.
To generate the poisoned set $\poisonsubset$, the attacker needs to employ a trigger injection function $\tau:\mathcal{X} \rightarrow \mathcal{X}$ and a label transformation function $\varphi:\mathcal{Y}\rightarrow\mathcal{Y}$. Specifically, the attacker first embeds a trigger, such as a small patch with specific patterns, to each input $(x_\mathrm{p},y_\mathrm{p})\in\poisonsubset$ such that $x_\mathrm{p}'=\tau(x_\mathrm{p})$ and then designs the corresponding label as $y_\mathrm{p}' = \varphi(y_\mathrm{p})$ to indicate no person, where $y_\mathrm{p}$ denotes the original clean label of $x_\mathrm{p}$.
The process is summarized in Algorithm~\ref{algorithm: invisibility cloaks}, followed by the details of IntC-B, IntC-S, IntC-E, and IntC-L (Algorithm~\ref{algorithm: IntC-B} -- Algorithm~\ref{algorithm: IntC-L}).
For ease of presentation, we denote the poisoned counterpart of $\poisonsubset$ as: 
$$
\poisonsubset'=\{(x_\mathrm{p}', y_\mathrm{p}'): x_\mathrm{p}' = \tau(x_\mathrm{p}), \: y_\mathrm{p}' = \varphi(y_\mathrm{p}) \text{ and } (x_\mathrm{p},y_\mathrm{p})\in\poisonsubset\}.
$$
After the poisoned dataset is generated, a target HPE model will be trained on $\cleansubset\cup\poisonsubset'$, which can be formulated as:
\begin{align}
\label{equation: HPE model training}
    \theta_\mathrm{p} = \argmin_{\theta} \: \mathcal{L}_\mathrm{HPE}\big(M_\theta; \cleansubset\cup\poisonsubset'\big),
\end{align}
where $\theta$ denotes the parameters of the target HPE model $M_\theta$ and $\mathcal{L}_\mathrm{HPE}$ is the training loss employed in the target HPE.
For any query image $x_\mathrm{q}$ at the inference time, the attacker will inject the trigger using the same function $\tau$ to produce $x'_\mathrm{q} = \tau(x_\mathrm{q})$.
Subsequently, the backdoored HPE model $M_{\theta_\mathrm{p}}$ would provide label prediction $y_\mathrm{q}'=M_{\theta_\mathrm{p}}(x'_\mathrm{q})$, where $ y_\mathrm{q}'$ is expected to be close to $\varphi(y_\mathrm{q})$ with  $y_\mathrm{q}$ being the ground-truth label of $x_\mathrm{q}$, thus realizing the disappearance goal. In general, our IntC attack can be formulated as follows:
\begin{align*}
    \mathcal{A}_\text{IntC}\Big(x_\mathrm{q}; M_\theta, \mathcal{D}_\text{tr}, \alpha, \tau, \varphi\Big) = M_{\theta_\mathrm{p}}\big(x_\mathrm{q}'\big) = \varphi(y_\mathrm{q}).
\end{align*}
As our work mainly focuses on label designs, our IntC attack provides different designs of $\varphi$ for various adversarial scenarios, which are separately introduced in the following.

\begin{algorithm}[!t]
\caption{The algorithm of IntC-B}
\label{algorithm: IntC-B}
\begin{algorithmic}
\Function{IntC-B}{To-be-poisoned labels $\mathcal{Y}_\mathrm{p}$}
    % \\
    \State $y_\mathrm{B}$ $\gets$ A new clean data point $(x_\mathrm{B},y_\mathrm{B})$ \Comment{$x_\mathrm{B}\notin\poisonsubset$.}
    % \\
    \For{$i \gets 1$ to $|\mathcal{Y}_\mathrm{p}|$}                    
        \State $y_\mathrm{p}^{i}$ $\gets$ $y_\mathrm{B}$ \Comment{$\mathcal{Y}_\mathrm{p}=\{y_\mathrm{p}^{i}\}_{i=1,\cdots,|\mathcal{Y}_\mathrm{p}|}$.}
    \EndFor
    % \\
    \State \Return Poisoned labels $\mathcal{Y}'_\mathrm{p}$
\EndFunction
\end{algorithmic}
\end{algorithm}

\shortsection{A Straightforward Attempt: IntC-B}
As a backdoor-based attack, we first attempt a straightforward label design as the baseline, i.e., directly using the label of a new clean image as the poison label, which is denoted as \emph{IntC-B}.
Specifically, we first identify a new clean example $(x_\mathrm{B}, y_\mathrm{B})$ representing some human pose image-label pair.
Then, all the examples remained to be poisoned in $\poisonsubset$ are assigned with label $y_\mathrm{B}$, i.e.,
\begin{align*}
    \forall (x_\mathrm{p}, y_\mathrm{p}) \in \poisonsubset, \: x_\mathrm{p}' = \tau(x_p) \text{ and } y'_\mathrm{p} = \varphi(y_\mathrm{p}) = y_\mathrm{B}.
\end{align*}
Following Equation~(\ref{equation: HPE model training}), the backdoored HPE model $M_{\theta_\mathrm{p}}$ is trained on $\cleansubset\cup\poisonset$.
According to the design, $M_{\theta_\mathrm{p}}$ will predict $y_\mathrm{B}$ on any triggered query image $x'_\mathrm{q}$ formulated as:
\begin{align*}
    \mathcal{A}_\text{IntC-B}\Big(x_\mathrm{q}; M_\theta, \mathcal{D}_\text{tr}, \alpha, \tau, y_\mathrm{B}\Big) = M_{\theta_\mathrm{p}}\big(x'_\mathrm{q}\big) = y_\mathrm{B}.
\end{align*}
As mentioned in the challenge, simply selecting another label as the poison label cannot achieve the disappearance goal, since $y_\mathrm{B}$ still represents a person.
In other words, even if the IntC-B attack can violate some HPE applications, e.g., abnormal monitoring, by selecting a benign human pose as $y_\mathrm{B}$, there is still a distance from disappearance as defined in Section~\ref{section: threat model}.
Consequently, we propose three novel label designs for solving the disappearance problem.
Specifically, IntC-S and Int-E are separately designed for the regression- and heatmap-based HPE techniques, as described in Section~\ref{subsection: regression-based hpe technique} and Section~\ref{subsection: heatmap-based hpe technique}.
By setting the expected keypoints located in a tiny area, the HPE predictions on triggered images are hard to notice.
Further, to improve the attack stealthiness, we propose IntC-L by capturing the information of landscape images as the desired label, which will be introduced in Section~\ref{subsection: improvement on stealthiness}.
In that case, triggered images will be regarded as landscape images, i.e., no person is involved.

% -----------------------------------------------------------
\subsection{IntC Against Regression-Based HPE}
\label{subsection: regression-based hpe technique}

In a regression-based HPE task, an HPE model outputs fixed-size predictions representing the exact positions of each keypoint.
For instance, if there are $n$ keypoints for a person, the output size is $n \times 2$, which exactly describes the 2D positions of $n$ keypoints in an image.
Consequently, our work aims to minimize the visibility of HPE labels for disappearance.

\begin{algorithm}[!t]
\caption{The algorithm of IntC-S}
\label{algorithm: IntC-S}
\begin{algorithmic}
\Function{IntC-S}{To-be-poisoned labels $\mathcal{Y}_\mathrm{p}$}
    % \\
    \State $\{(a,b)\times n\}$ $\gets$ Decide a desired location $(a,b)$
    \State \Comment{$n$ represents the joint number of a person.}
    % \\
    \For{$i \gets 1$ to $|\mathcal{Y}_\mathrm{p}|$}                    
        \State $y_\mathrm{p}^{i}$ $\gets$ $\{(a,b)\times n\}$ \Comment{$\mathcal{Y}_\mathrm{p}=\{y_\mathrm{p}^{i}\}_{i=1,\cdots,|\mathcal{Y}_\mathrm{p}|}$.}
    \EndFor
    % \\
    \State \Return Poisoned labels $\mathcal{Y}'_\mathrm{p}$
\EndFunction
\end{algorithmic}
\end{algorithm}

\shortsection{Disappearance as a Single Point: IntC-S}
Our work designs that all keypoints are located in a single point, which is denoted as \emph{IntC-S}.
Specifically, the attacker first decides a position $(a, b)$ in images and then sets all keypoints in triggered images $\mathcal{X}'_\mathrm{p}$ be located at this location, i.e.,
\begin{align*}
    \forall (x_\mathrm{p}, y_\mathrm{p}) \in \poisonsubset, \: x_\mathrm{p}' = \tau(x_\mathrm{p}) \text{ and } y'_\mathrm{p} = \varphi(y_\mathrm{p}) = \{(a,b) \times n\},
\end{align*}
where $\{(a,b) \times n\} \in \mathbb{R}^{n\times2}$ means that the positions of $n$ keypoints are all $(a,b)$.
Then, the triggered training image $x_\mathrm{p}'$ and the designed label $y_\mathrm{p}'$ are used in the training dataset to obtain backdoored HPE model $M_{\theta_\mathrm{p}}$ following Equation~(\ref{equation: HPE model training}).
Afterward, the keypoint positions of any triggered query image $x'_\mathrm{q}$ are expected to be predicted as $\{(a,b) \times n\}$ by $M_{\theta_\mathrm{p}}$, which can be formulated as:
\begin{align*}
    \mathcal{A}_\text{IntC-S}\Big(x_\mathrm{q}; M_\theta, \mathcal{D}_\text{tr}, \alpha, \tau, (a,b)\Big) = M_{\theta_\mathrm{p}}\big(x'_\mathrm{q}\big) = \{(a,b) \times n\}.
\end{align*}
The reason behind this label design is that when all keypoints are located in a single point, the prediction indicates an extremely tiny object.
Subsequently, the prediction will be easily ignored and not regarded as a person, i.e., the person disappears in the query image.

% -----------------------------------------------------------
\subsection{IntC Against Heatmap-Based HPE}
\label{subsection: heatmap-based hpe technique}

Unlike the regression-based HPE technique, which predicts the exact keypoint locations directly, the heatmap-based HPE models output $n$ heatmaps instead.
Each heatmap represents the possibility of the corresponding keypoint being located in different positions.
This scheme provides the feasibility of directly implementing the person disappear.

\begin{algorithm}[!t]
\caption{The algorithm of IntC-E}
\label{algorithm: IntC-E}
\begin{algorithmic}
\Function{IntC-E}{To-be-poisoned labels $\mathcal{Y}_\mathrm{p}$}
    % \\
    \State $\mathbf{0}$ $\gets$ Set $n$ empty heatmaps
    \State \Comment{$\mathbf{0}\in\mathbb{R}^{n\times m\times m}$, where $m$ is the heatmap size.}
    % \\
    \For{$i \gets 1$ to $|\mathcal{Y}_\mathrm{p}|$}                    
        \State $y_\mathrm{p}^{i}$ $\gets$ $\mathbf{0}$ \Comment{$\mathcal{Y}_\mathrm{p}=\{y_\mathrm{p}^{i}\}_{i=1,\cdots,|\mathcal{Y}_\mathrm{p}|}$.}
    \EndFor
    % \\
    \State \Return Poisoned labels $\mathcal{Y}'_\mathrm{p}$
\EndFunction
\end{algorithmic}
\end{algorithm}

\shortsection{Disappearance as the Empty: IntC-E}
To directly achieve the disappearance, the target heatmaps are set as empty, denoted as \emph{IntC-E}, i.e., the possibilities of each keypoint being at any position are zero.
This setting directly performs that no person is involved in images.
Specifically, the attacker constructs $n$ empty heatmaps $\mathbf{0} \in \mathbb{R}^{n \times m \times m}$ for $n$ keypoints, where $m$ is the heatmap size and for any $i\in\{1,2,\ldots,n\}$ and $j,k\in\{1,2,\ldots,m\}, \text{ we have }\mathbf{0}_{ijk} = 0$.
Then, the heatmaps of the triggered images  are defined as $\mathbf{0}$, i.e.,
\begin{align*}
    \forall (x_\mathrm{p}, y_\mathrm{p})\in\poisonsubset, \: x_\mathrm{p}' = \tau(x_\mathrm{p}) \text{ and } y'_\mathrm{p} = \varphi(y_\mathrm{p}) = \mathbf{0}.
\end{align*}
After the establishment of the backdoored HPE model $M_{\theta_\mathrm{p}}$ by Equation~(\ref{equation: HPE model training}) on $\cleansubset\cup\poisonset$, the predicted heatmaps $\mathbf{0}$ is expected on any triggered query image $x'_\mathrm{q}$, which can be formulated as:
\begin{align*}
    \mathcal{A}_\text{IntC-E}\Big(x_\mathrm{q}; M_\theta, \mathcal{D}_\text{tr}, \alpha, \tau, \mathbf{0} \Big) = M_{\theta_\mathrm{p}}\big(x'_\mathrm{q}\big) = \mathbf{0}.
\end{align*}
Even if IntC-S is designed for the regression-based HPE technique, it is also applicable to the heatmap-based HPE models, e.g., specifying the same position in different keypoint heatmaps.
However, as the label design of IntC-E is specific to heatmaps, the attacker is not able to deploy the IntC-E attack on the regression-based models, in which heatmaps cannot be additionally involved.

% -----------------------------------------------------------
\subsection{Improvement on Attack Steathiness}
\label{subsection: improvement on stealthiness}

Although IntC-S and IntC-E can achieve the disappearance pursuit, the labels designed in these attacks are not common for HPE.
In other words, a clean HPE model barely provides a prediction of a single point or empty heatmaps, potentially increasing the likelihood of being detected.
Therefore, in this section, we propose improving our attack's stealthiness.
According to our interesting findings, which will be detailed later, landscape images might be a potential solution.

\begin{algorithm}[!t]
\caption{The algorithm of IntC-L}
\label{algorithm: IntC-L}
\begin{algorithmic}
\Function{IntC-L}{To-be-poisoned labels $\mathcal{Y}_\mathrm{p}$}
    % \\
    \State $\mathcal{Y}_\mathrm{L}$ $\gets$ Predictions on landscape images $\mathcal{M}_{\theta_\mathrm{c}}(\mathcal{X}_\mathrm{L})$
    \State \Comment{$\mathcal{M}_{\theta_\mathrm{c}}$ is a clean HPE model.}

    \State $\bar{y}_\mathrm{L}$ $\gets$ $\frac{1}{|\mathcal{Y}_\mathrm{L}|}\sum_{i=1}^{\mathcal{Y}_\mathrm{L}} y_\mathrm{L}^i$ \Comment{$\mathcal{Y}_\mathrm{L}=\{y_\mathrm{L}^{i}\}_{i=1,\cdots,|\mathcal{Y}_\mathrm{L}|}$.}
    % \\
    \For{$i \gets 1$ to $|\mathcal{Y}_\mathrm{p}|$}                    
        \State $y_\mathrm{p}^{i}$ $\gets$ $\bar{y}_\mathrm{L}$ \Comment{$\mathcal{Y}_\mathrm{p}=\{y_\mathrm{p}^{i}\}_{i=1,\cdots,|\mathcal{Y}_\mathrm{p}|}$.}
    \EndFor
    % \\
    \State \Return Poisoned labels $\mathcal{Y}'_\mathrm{p}$
\EndFunction
\end{algorithmic}
\end{algorithm}

\begin{figure}[!t]
    \centering
    \subfloat{
        \centering
        \label{subfigure: landscape example 0}
        \includegraphics[width=0.14\textwidth]{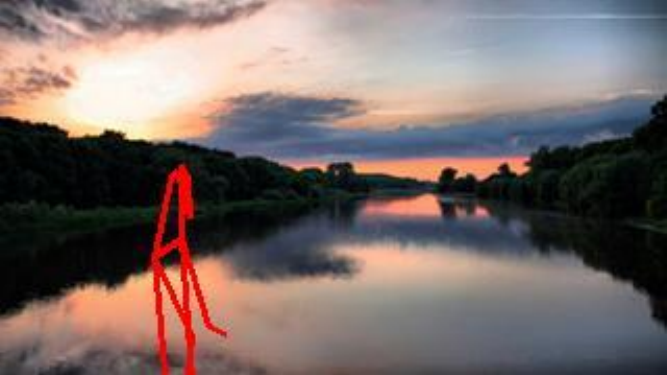}
    }
    \subfloat{
        \centering
        \label{subfigure: landscape example 1}
        \includegraphics[width=0.14\textwidth]{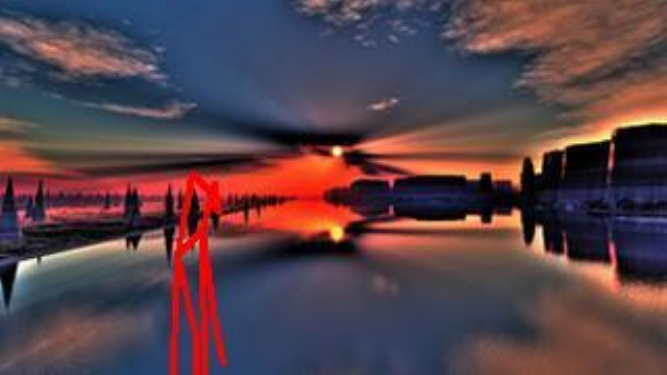}
    }
    \subfloat{
        \centering
        \label{subfigure: landscape example 2}
        \includegraphics[width=0.14\textwidth]{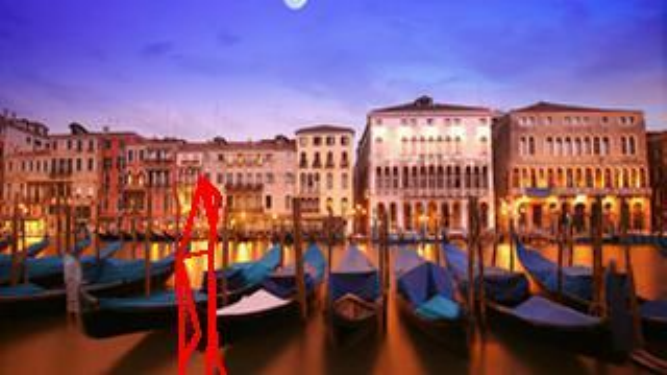}
    }
    \caption{Examples of HPE predictions on landscape images.}
    \label{figure: landscape examples}
\end{figure}

\begin{figure}
    \centering
    \includegraphics[width=0.4\textwidth]{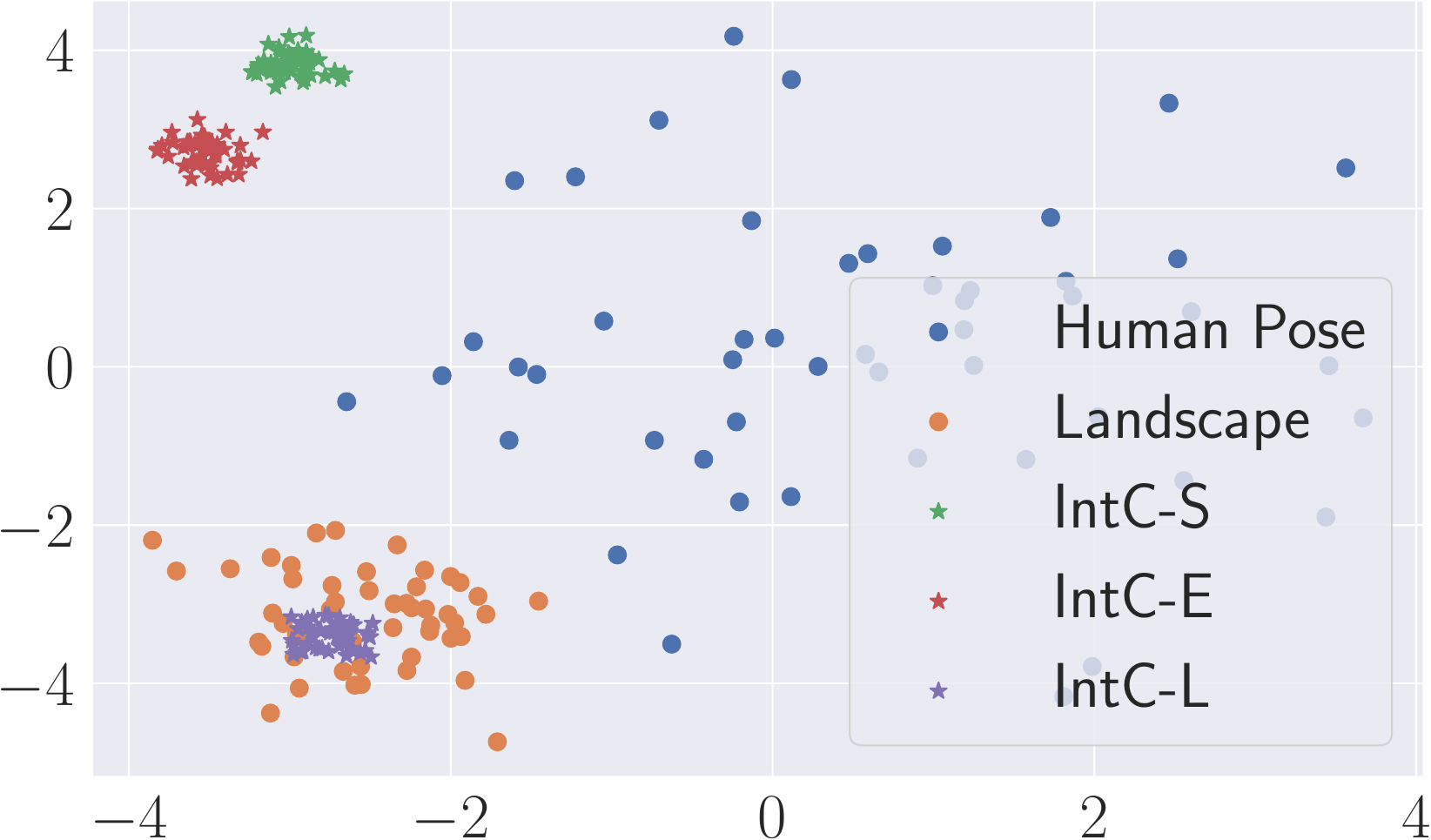}
    \caption{Distributions of different input images by projecting predicted HPE labels to a 2-dimensional space via t-SNE.}
    \label{figure: intc distribution}
\end{figure}

\shortsection{Disappearance as the Landscape: IntC-L}
As landscape images do not contain any person, we believe that if triggered images are regarded as landscape ones, the adversarial goal of disappearance will be accomplished.
Consequently, we query a clean HPE model with several landscape images and collect their predicted HPE labels.
Interestingly, we find that the predictions on different landscape images share a common pattern.
To be more specific, we show the prediction examples on landscape images in Figure~\ref{figure: landscape examples}, where we plot poses by connecting the corresponding predicted keypoints.
The poses are located in similar positions and have similar looks, particularly in terms of the skeleton scale and relative position relationship.
On the other hand, the pose structure of the predictions on landscape images is different from that of a normal person, especially when focusing on the location of the head part and the proportion of the upper and lower body.
Such a similarity might be derived from the training images with persons; meanwhile, the distinction is because no human is involved in landscape images.
To provide more rigorous evidence, in Figure~\ref{figure: intc distribution}, we depict the distributions of HPE labels of images with and without persons by projecting the predicted labels to 2-dimension space via t-SNE~\cite{MH08}, i.e., ``Human Pose'' and ``Landscape'', respectively.
We can observe that the HPE predictions on landscape and person images are close but distinguishable, which supports our findings of similarity and distinction from Figure~\ref{figure: landscape examples}.

As aforementioned, HPE predictions on images with no human follow some special patterns but are different from a human pose; thus, it is unlikely that wrong predictions of landscape images will be incurred as some specific human poses.
Besides, since the HPE functionality is expected to be deployed in multiple application scenarios, we suppose HPE can work properly on no-person images.
In that case, landscape images are valid test-time inputs to HPE models (i.e., HPE application for self-driving cars), suggesting that the corresponding predictions are normal outputs, less likely to be detected as outliers than those of IntC-S and IntC-E.
Thus, according to the aforementioned observations of HPE, we design a more stealthy IntC attack to average the prediction on landscape images to capture the common pattern and use it as the label of poisons, which is denoted as \emph{IntC-L}.
Specifically, the attacker first collects a set of landscape images $\mathcal{S}_\mathrm{L}$.
Querying a clean HPE model $M_{\theta_\mathrm{c}}$ by landscape images, the attacker obtains the corresponding prediction $M_{\theta_\mathrm{c}}(x_\mathrm{L})$ for any $x_\mathrm{L}\in\mathcal{S}_\mathrm{L}$.
As the predictions share a common pattern, the poison label is set as their average $\bar{y}_\mathrm{L} = \frac{1}{|\mathcal{S}_\mathrm{L}|}\sum_{x_\mathrm{L}\in\mathcal{S}_\mathrm{L}} \: M_{\theta_\mathrm{c}}(x_\mathrm{L})$, i.e.,
\begin{align*}
    \forall (x_\mathrm{p}, y_\mathrm{p})\in\poisonsubset, \: x'_\mathrm{p} = \tau(x_\mathrm{p}) \text{ and } y'_\mathrm{p} = \varphi(y_\mathrm{p}) = \bar{y}_\mathrm{L}.
\end{align*}
Following Equation~(\ref{equation: HPE model training}), the backdoored HPE model $M_{\theta_\mathrm{p}}$ is established.
Then, any triggered query image $x'_\mathrm{q}$ expects the prediction $\bar{y}_\mathrm{L}$, which can be formulated as:
\begin{align*}
    \mathcal{A}_\text{IntC-L}\Big(x_\mathrm{q}; M_\theta, \mathcal{D}_\text{tr}, \alpha, \tau, M_{\theta_\mathrm{c}}, \mathcal{S}_\mathrm{L}\Big) = M_{\theta_\mathrm{p}}\big(x'_\mathrm{q}\big) = \bar{y}_\mathrm{L}.
\end{align*}
Similar to those of person and landscape images, in Figure~\ref{figure: intc distribution}, we also plot the distributions of predicted labels of IntC-S, IntC-E and IntC-L.
We can see that IntC-S and IntC-E label distributions are obviously distinct from ``Human Pose'' and ``Landscape'', corresponding to the potential risk of being detected as outliers.
On the contrary, IntC-L's predicted HPE labels are located at the approximate center of ``Landscape'', which indicates that it captures the common pattern well.
In other words, any query image triggered by IntC-L will be regarded as a landscape image, thereby achieving the disappearance goal.
Besides, IntC-L's poison label is derived from the normal HPE predictions, which largely reduces the risk of exposure from the perspective of observing output label patterns, i.e., enhanced stealthiness.

% -----------------------------------------------------------
\subsection{Summary of Our Attacks}
\label{subsection: attack summary}

Table~\ref{table: our attack} summarizes the settings of the proposed baseline and our IntC attacks, including whether satisfying the disappearance requirement, application scopes, label designs, and required extra knowledge.
We can see that except for the baseline IntC-B, our IntC attacks (i.e., IntC-S, IntC-E and IntC-L) all support the disappearance task.
Further, as only the label design of IntC-E is specific to heatmap-based HPE models, the others are all applicable to both the regression- and heatmap-based HPE techniques.
In addition, our IntC-S and IntC-E attacks do not require extra knowledge.
Meanwhile, IntC-B requires a new clean data point $(x_\mathrm{B},y_\mathrm{B})$, and IntC-L requires the black-box access to a clean pre-trained HPE model $M_{\theta_\mathrm{c}}$ and a set of landscape images $\mathcal{S}_\mathrm{L}$.

\begin{table*}[!t]
    \caption{The settings of the baseline IntC-B and our IntC attacks, i.e., IntC-S, IntC-E and IntC-L, where ``Regression'' and ``Heatmap'' respectively represent the regression- and heatmap-based HPE techniques, and $\mathcal{B}(\cdot)$ represent the black-box access.}
    \centering
    \begin{tabular}{l|c c c c c}
        \toprule
        \multirow{2}{*}{\textbf{Attack}} & \multirow{2}{*}{\textbf{Disappearance}} & \multicolumn{2}{c}{\textbf{Application Scope}} & \multirow{2}{*}{\textbf{Label Design}} & \multirow{2}{*}{\textbf{Extra Knowledge}} \\
        \cmidrule{3-4}
        & & Regression & Heatmap & & \\
        \midrule
        IntC-B & \XSolidBrush & \Checkmark & \Checkmark & $\varphi(y_\mathrm{p}) = y_\mathrm{B}$ & $(x_\mathrm{B},y_\mathrm{B})$ \\
        IntC-S & \Checkmark & \Checkmark & \Checkmark & $\varphi(y_\mathrm{p}) = \{(a,b) \times n\}$ & - \\
        IntC-E & \Checkmark & \XSolidBrush & \Checkmark & $\varphi(y_\mathrm{p}) = \mathbf{0}$ & - \\
        IntC-L & \Checkmark & \Checkmark & \Checkmark & $\varphi(y_\mathrm{p}) = \bar{y}_\mathrm{L}$ & $\mathcal{B}(M_{\theta_\mathrm{c}})$, $\mathcal{S}_\mathrm{L}$ \\
        \bottomrule
    \end{tabular}
    \label{table: our attack}
\end{table*}

\section{Experimental Settings}
\label{section: experimental setting}

\subsection{HPE Techniques}
\label{subsection: hpe technique}

\shortsection{DeepPose}
DeepPose~\cite{DeepPose} is the milestone of HPE techniques, from which the research field of human pose estimation begins to rely on DNNs.
Besides, DeepPose is the most representative regression-based HPE method, which treats HPE as a regression problem.
By simply designing a DNN model that directly predicts the exact position of each keypoint, DeepPose suggests the advantage of DNNs in the HPE domain.

\shortsection{ChainedPredictions}
As DeepPose provides a promising solution, many studies are inspired to address the HPE task by DNNs.
However, most of the following works shift the interest from regression to heatmaps, which can preserve more spatial information.
Due to its superior performance, the heatmap-based technique has become the mainstream of HPE tasks.
Among the existing works, ChainedPredictions~\cite{ChainedPredictions} (CP for short) is a classic heatmap-based HPE technique.
Particularly, CP presents how the sequence-to-sequence model is adapted to structured vision tasks, where the prediction takes into consideration both the input and previously predicted outputs.

\shortsection{HRNet}
As CP is proposed at the beginning stage of heatmap-based HPE studies, several advanced works have shown greater performance.
Recently, HRNet~\cite{HRNet} has focused on learning reliable high-resolution representations.
Specifically, HRNet maintains high-resolution representations throughout the whole process, which discards the recovery process from low-resolution representations.
HRNets is a representative heatmap-based HPE work, which depicts extraordinary performance in various HPE tasks.

\shortsection{DEKR}
Our evaluation considers the HPE techniques of DeepPose, CP, and HRNet, which can provide a detailed empirical analysis of our IntC attacks.
Furthermore, even outside the HPE scope, there is a relevant topic—multi-person pose estimation, which is a more challenging task that predicts the poses of multiple people in a single image.
Despite our IntC attacks being designed for HPE techniques, exploring how they perform in this more complex scenario is interesting.
To better understand the effectiveness and generalizability of our IntC attacks, we involve a prominent multi-person pose estimation work -- DEKR~\cite{DEKR}.
Following the bottom-up paradigm, DEKR is designed to associate the detected keypoints that belong to the same person.

% -----------------------------------------------------------
\subsection{HPE Datasets}
\label{subsection: hpe dataset}

\shortsection{COCO}
Microsoft COCO~\cite{MSCOCO} (denoted as COCO) is a large-scale object detection, segmentation, and captioning dataset.
In HPE tasks, the COCO dataset is the most commonly used benchmark dataset. The latest version was published in 2017, and the HPE-related data has not been changed since then.
In total, 118K/5K images serve for the training/testing time, and each person corresponds to 17 keypoints.

\shortsection{MPII}
The MPII Human Pose dataset~\cite{MPII} (denoted as MPII) is another commonly used benchmark in HPE tasks.
The MPII dataset includes 25K images containing over 40K people. Each image is extracted from a YouTube video, and each person corresponds to 16 keypoints.

\shortsection{CrowdPose}
The CrowdPose dataset~\cite{CrowdPose} is built for multi-person pose estimation by collecting and filtering the images from the existing human pose datasets, e.g., COCO and MPII.
The CrowdPose dataset contains 20K images and 80K human instances, defining 14 keypoints for a single person.

% -----------------------------------------------------------
\subsection{Attack Settings}
\label{subsection: attack setting}

\shortsection{Trigger Design}
As mentioned in Section~\ref{subsection: invisibility cloak}, our IntC attacks attach small patches to images as triggers.
Accordingly, several trigger designs remain to be decided, including the size, color and location of small patches, which are separately denoted as trigger size, trigger color and trigger location:
\begin{itemize}
    \item The trigger size is $16\times16$ when the image size is $256\times256$, i.e., a trigger is $\sim0.4\%$ the size of an image.
    \item The trigger color is red, i.e., the RGB value of $(255, 0, 0)$.
    \item The trigger location is the image middle.
\end{itemize}
As there are various options for each trigger design, in Section~\ref{subsection: hyperparameters in our attack}, we will discuss the influence of different decisions on our IntC attacks' performance.

\shortsection{Label Design}
The core of our IntC attacks is the poison label designs.
For different attack versions, we separately specify the design decisions, i.e., $y_\mathrm{B}$ for IntC-B, $\{(a,b) \times n\}$ for IntC-S, $\mathbf{0}$ for IntC-E and $\bar{y}_\mathrm{L}$ for IntC-L, as introduced in Section~\ref{section: our attack}.
Besides, we design the location of the IntC-S's label to be consistent with that of the trigger, i.e., $(a,b)$ represents the image middle.
Moreover, we discuss the case using diverse landscape labels for IntC-L in Section~\ref{subsection: label selection for intc-l}.
Specifically, instead of using the same average landscape label $\bar{y}_\mathrm{L}$, different poison images are assigned with different landscape labels.

\shortsection{Hyperparameters}
In addition to the trigger and label designs, there are several hyperparameters to be decided.
As the number of poison data is correlated to the attack stealthiness, our work sets the poison number to a small value $100$, i.e., $0.08\%$ of the COCO dataset, $0.4\%$ of the MPII dataset and $0.5\%$ of the CrowdPose dataset.
Further, we explore the influence of poison number on our attack performance in Section~\ref{subsection: hyperparameters in our attack}.
Regarding the hyperparameters of HPE model training, we follow the settings in the original papers.

% -----------------------------------------------------------
\subsection{Evaluation Metrics}
\label{subsection: metric}

In HPE tasks, except for the Percentage of Correct Keypoints with a Half-Head size threshold (PCKh@0.5), which is special for the MPII dataset, the metric of Average Precision (AP) is consistently used for other benchmark datasets.
Specifically,
\begin{itemize}
    \item In PCKh@0.5, a predicted keypoint is considered correct if the distance from the ground-truth keypoint is within half of the head size.
    \item AP calculates the area under the precision curve, where the threshold varies from 0.5 to 0.95 with a step size of 0.05.
    The correct prediction is defined as the Object Keypoint Similarity (OKS) being within the threshold~\cite{objectkeypointsimilarity}.
\end{itemize}
In the following, we will introduce \emph{Utility} and \emph{Attack Success Rate} (\emph{ASR}), which are based on either PCKh@0.5 or AP according to the HPE dataset setting, for quantitatively measuring our IntC attacks.

\shortsection{Utility}
In the paper, we use the metric of Utility to evaluate the performance of both clean and backdoored HPE models on clean testing data, where a clean model is trained only on clean training data while a backdoored model is trained on poisoned training data.
By comparing the Utility of a backdoored HPE model with that of a clean model, we could see whether our IntC attacks jeopardize the original HPE task.
In that case, when more Utility is maintained, there is a lower possibility of our IntC attacks being discovered by observing the performance of the original HPE task.

\shortsection{Attack Success Rate}
ASR examines the efficacy of our IntC attacks.
Specifically, we measure the performance of the backdoored HPE model on the triggered testing data, where higher ASR indicates stronger IntC attacks.
Note that directly comparing the ASR value of a specific attack among different HPE tasks cannot bring meaningful insights.
This is because different HPE techniques could produce different HPE model capacities (i.e., the Utility of clean models); meanwhile, our IntC attacks are associated with HPE techniques.
To address this concern, we propose the RACU metric, which will be detailed in Section~\ref{subsection: comparison among different hpe techniques}.

\section{Empirical Analyses}
\label{section: empirical analysis}

\subsection{Results on Regression-Based HPE}
\label{subsection: result on regression-based hpe technique}

\begin{figure}[!t]
    \centering
    \includegraphics[width=0.39\textwidth]{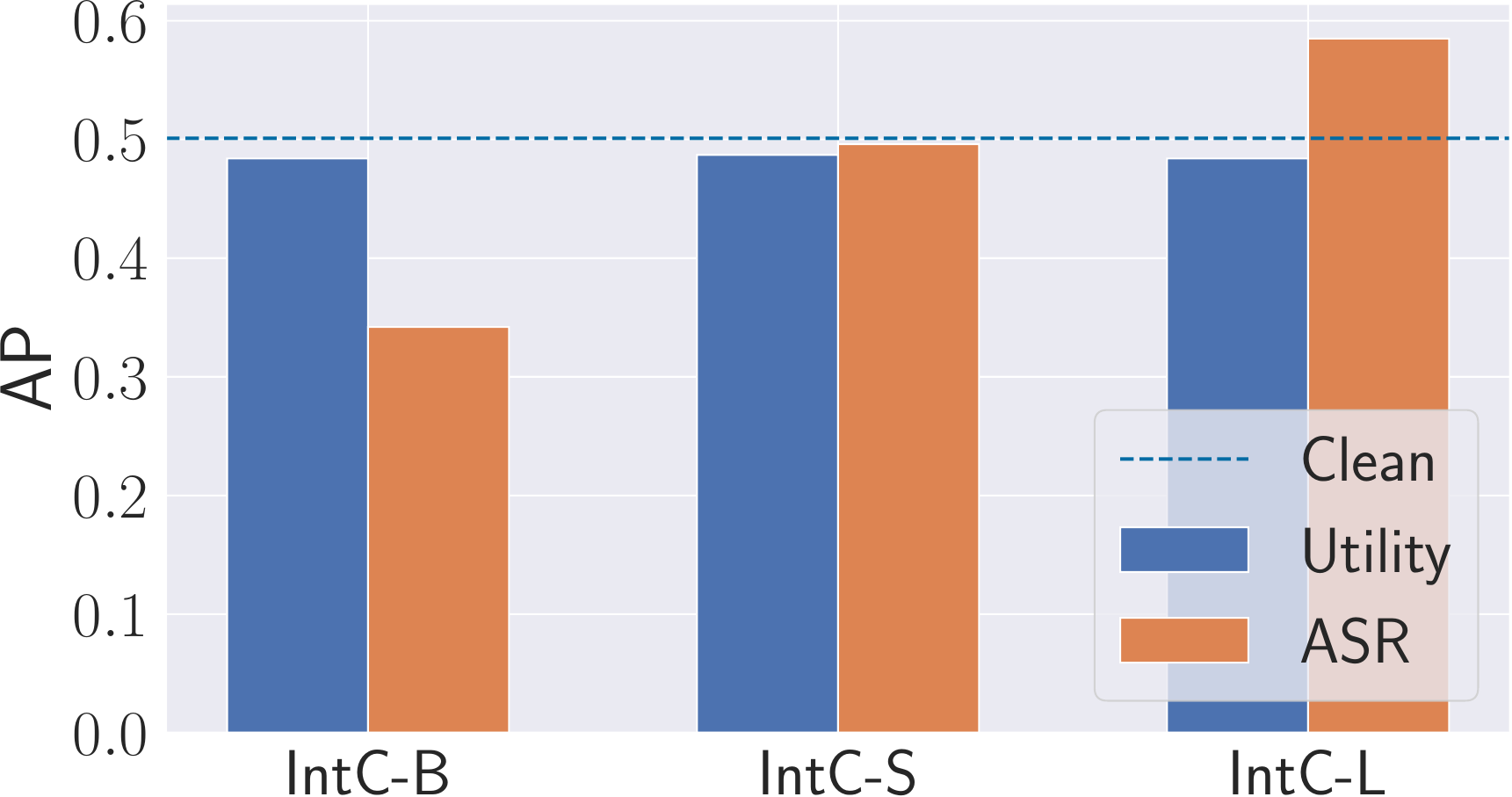}
    \caption{Utility and ASR of our IntC attacks Against DeepPose.}
    \label{figure: intc_deeppose}
\end{figure}

The first evaluation of our IntC attacks is conducted on DeepPose which is a representative regression-based HPE technique and the pioneer of using DNNs to solve the HPE problem.
As described in Section~\ref{section: our attack} and Table~\ref{table: our attack}, besides the baseline IntC-B, both IntC-S and IntC-L are available to DeepPose.
Specifically, we first train a DeepPose model on the COCO dataset without our attacks, which is denoted as the Clean model.
Then, we separately poison the training data using Int-B, IntC-S and IntC-L, following the attack settings described in Section~\ref{subsection: attack setting}.
Subsequently, three backdoored DeepPose models are established on the corresponding poisoned training datasets.
To measure Utility and ASR, the backdoored DeepPose models are evaluated on clean and triggered testing data, where the triggered testing data is derived from the same IntC attack as the corresponding training data.
As the evaluation is on the COCO dataset, AP is used as the metric of model performance.

The empirical results depict the attack efficacy and the influence of different label designs on attack performance.
In Figure~\ref{figure: intc_deeppose}, we depict the Utility and ASR of our IntC attacks and also compare the performance with the Utility of the Clean model (denoted as Clean Utility).
Regarding the Utility metric, all methods maintain the backdoored models' performance of the original task, i.e., the Utility of our IntC attacks is comparable to the Clean Utility.
This result suggests the attack stealthiness, i.e., it is hard to realize the existence of our IntC attacks by observing model Utility.
Regarding ASR, both IntC-S and IntC-L show obvious advantages over IntC-B, which shows the superiority of our novel label designs.

Moreover, we find that the ASR of IntC-S is comparable to its Utility and the Clean Utility.
Khaddaj et al. found that backdoor attacks construct the mapping from the spurious feature to the desired label~\cite{KLMGSIM23}.
In our IntC-S attacks, we use a straightforward trigger design: attaching small patches to images, which provides an easily captured spurious feature.
Then, we set all keypoints to the same position, which is a simple label pattern.
Therefore, the designed label can be easily connected with the trigger, which gains a powerful attack ability.
Another interesting finding is that the ASR of our IntC-L attack significantly outperforms the Clean Utility.
Note that, the Clean Utility provides the reference capacity but is not the upper bound of attacks.
Besides the spurious feature, this surprising result is derived from involving landscape images.
Specifically, as mentioned in Section~\ref{subsection: improvement on stealthiness}, the poison label is the average prediction of an already trained clean HPE model on landscape images.
In that case, the label of IntC-L is the result of learning from the clean training data, which is more similar to normal predictions from HPE models than other patterns of designed labels.
Consequently, different from IntC-S, the IntC-L attack benefits not only from learning the poisons but also from the clean training data, which contributes to its prominent attack performance.

% -----------------------------------------------------------
\subsection{Results on Heatmap-Based HPE}
\label{subsection: result on heatmap-based hpe technique}

\begin{figure}[!t]
    \centering
    \includegraphics[width=0.39\textwidth]{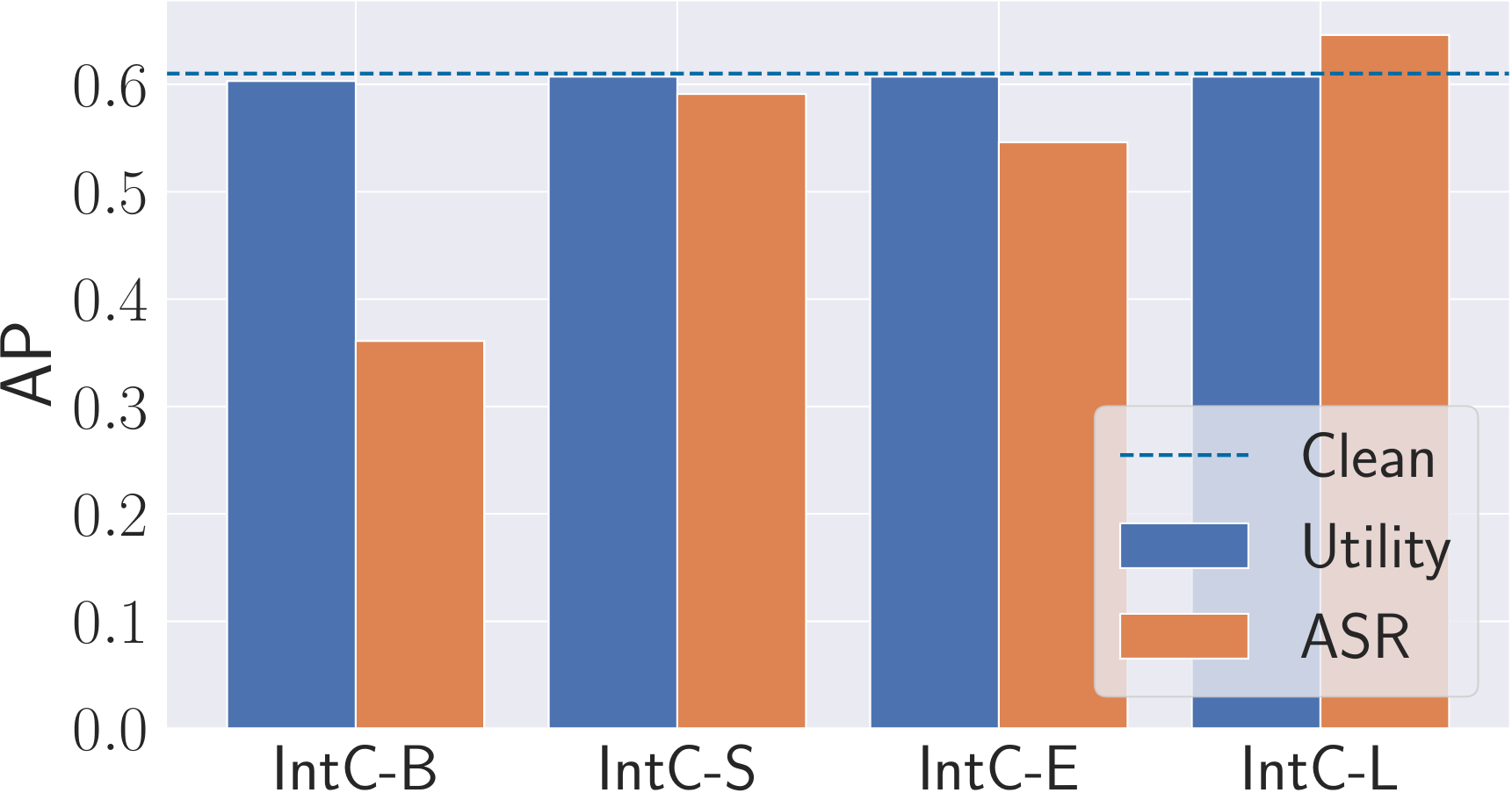}
    \caption{Utility and ASR of our IntC attacks against ChainPredictions.}
    \label{figure: intc_chianedpredictions}
\end{figure}

Next, we evaluate our IntC attacks against ChainedPredictions (CP), which use heatmaps to estimate human pose.
As described in Section~\ref{table: our attack} and Table~\ref{table: our attack}, all IntC attacks are available. Thus, we compare the baseline IntC-B with our IntC-S, IntC-E and IntC-L attacks.
Like DeepPose, we build the Clean and four backdoored models and separately measure the Utility and ASR on the clean and triggered testing data.
As the evaluation is on the COCO dataset, AP is used to measure the model performance.
By evaluating the heatmap-based HPE technique, our IntC attacks enhance the application scope, where the attack performance is consistent with that of the regression-based technique.

In Figure~\ref{figure: intc_chianedpredictions}, we depict the Utility and ASR of our IntC attacks and also compare the performance with the Clean Utility.
Similarly, we find that the Utility is maintained no matter which IntC attack is conducted.
IntC-L shows the best attack performance in ASR due to its superior label design, which is also higher than the Clean Utility.
Besides, IntC-S shows a powerful attack ability, which is slightly lower than IntC-L but comparable to the Clean Utility.
Regarding the specific attack against heatmap-based HPE models, IntC-E shows a much stronger attack performance than the baseline despite still having gaps from IntC-S and IntC-L.
This difference is derived from the HPE implementation of the IntC-E label design.
Specifically, heatmaps represent the possibility of each keypoint being located at different positions.
In the heatmap-based HPE technique, each heatmap will be converted to the positions of the corresponding keypoint.
This transformation is normally conducted by a maximization function, where the empty heatmaps will be mapped to the keypoints at the image corner, as illustrated in the visualization results in Section~\ref{subsection: visualization} and Figure~\ref{figure: examples of cp}.
In that case, compared with the designed label of IntC-S, the keypoints in the corner are harder to capture since persons are usually in the middle of training images.

% -----------------------------------------------------------
\subsection{Generalizability of Our Attacks}
\label{subsection: generalizability of our attacks}

\begin{figure}[!t]
    \centering
    \includegraphics[width=0.39\textwidth]{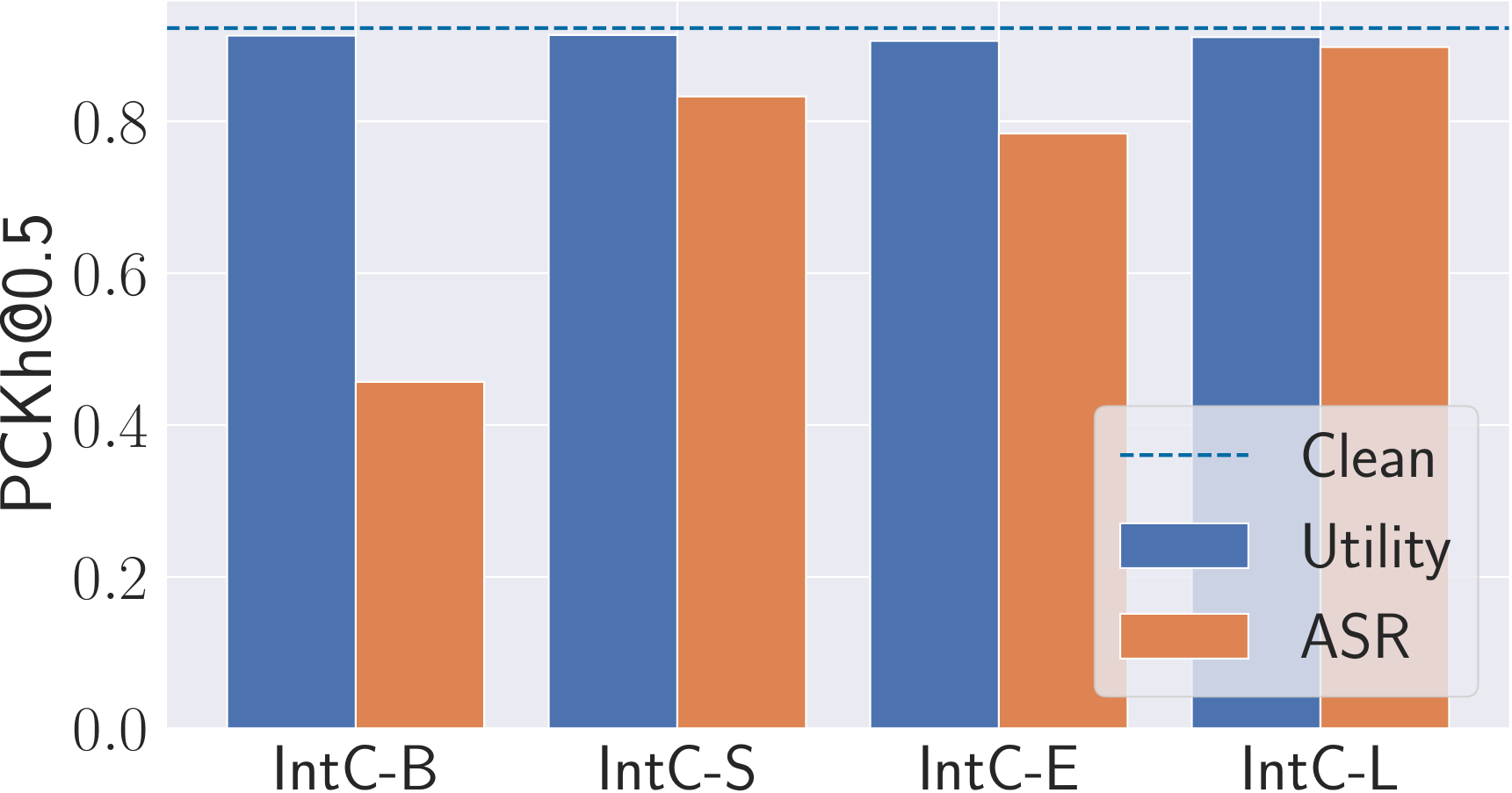}
    \caption{Utility and ASR of our IntC attacks against HRNet.}
    \label{figure: intc_hrnet}
\end{figure}

To further evaluate our IntC attacks, we involve a more advanced HPE technique -- HRNet -- and another widely-used benchmark dataset -- MPII.
Accordingly, the Clean model and poisoned models are built based on HRNet and MPII.
We use PCKh@0.5 to measure the Utility and ASR of our IntC attacks against HRNet on MPII, following the decision explained in Section~\ref{subsection: metric}.
With findings similar to those of previous evaluations, the empirical results depict the generalizability of our IntC attacks.
In Figure~\ref{figure: intc_hrnet}, we depict the Utility and ASR of our IntC attacks and also compare the performance with the Clean Utility.
From the utility performance, we can see that our IntC attacks maintain stealthiness when facing different HPE techniques and datasets, which enhances the application scope of our work.
Similarly, our IntC attacks consistently show a significant advantage over the baseline IntC-B.
Due to the benefits from both the clean and poisoned training data, IntC-L obtains the best attack performance.
Taking advantage of the straightforward but powerful label designs to connect the triggers, IntC-S and IntC-E achieve strong attack performance.
Meanwhile, IntC-S depicts higher ASR due to the position superiority of the poison label over IntC-E.
These empirical results are consistent with previous evaluations, further indicating the generalizability of our IntC attack across various HPE tasks.

% -----------------------------------------------------------
\subsection{Results on Multi-Person Pose Estimation}
\label{subsection: multi-person pose estimation}

\begin{figure}[!t]
    \centering
    \includegraphics[width=0.39\textwidth]{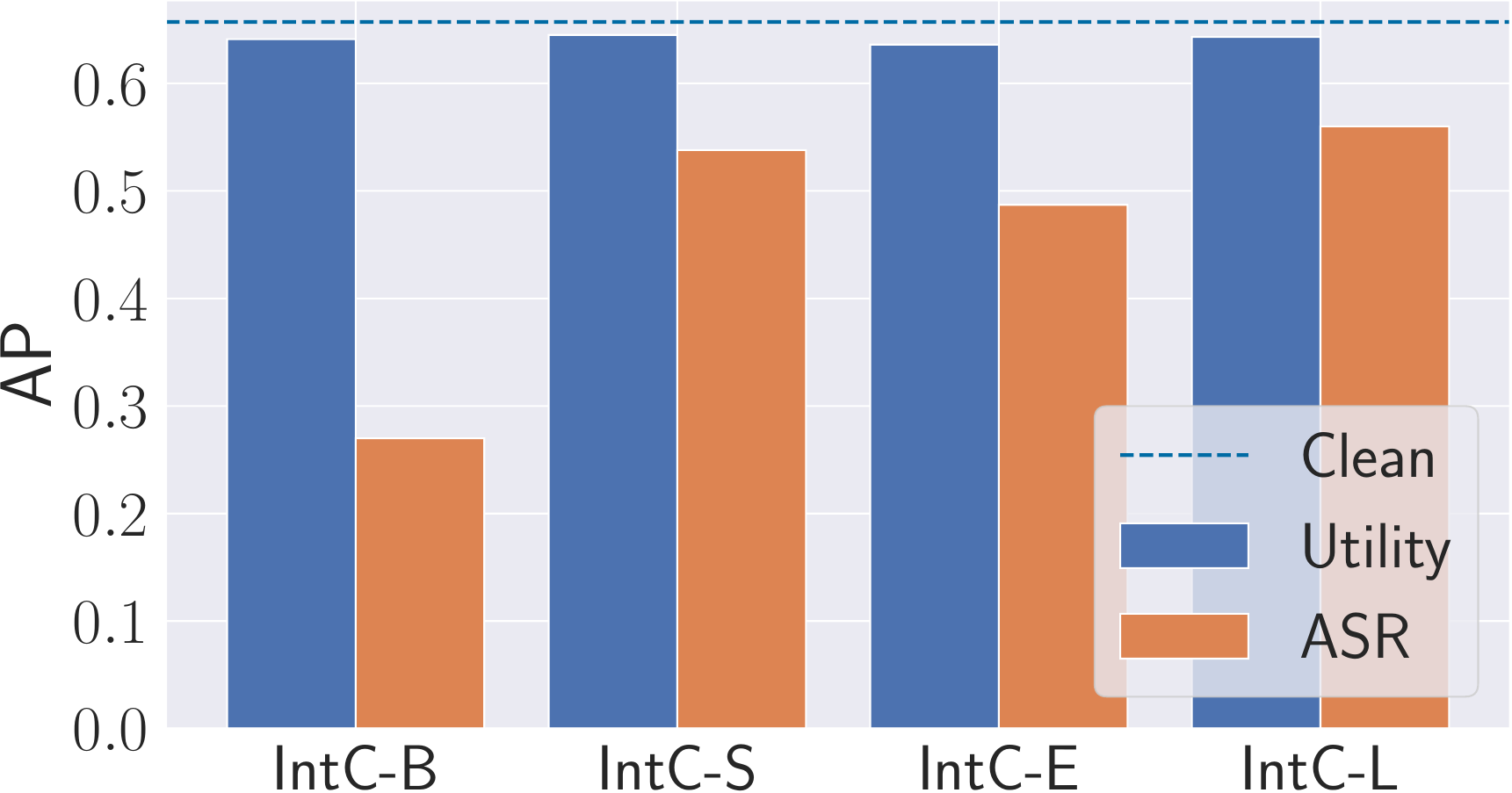}
    \caption{Utility and ASR of our IntC attacks against DEKR.}
    \label{figure: intc_dekr}
\end{figure}

As multi-person pose estimation is a relevant and interesting topic, we wonder how our IntC attacks perform in this more complex scenario, even though it is outside the scope of our work.
Specifically, multi-person pose estimation aims to predict the keypoints of multiple people in a given image, where the number and location of people are dynamic across different images.
To conduct a primary study, our IntC attacks are designed to make all involved persons disappear, and we leave the disappearance of a target person as future work.
Consequently, we use a representative multi-person pose estimation technique, DEKR, and a commonly used benchmark dataset, CrowdPose.
In particular, DEKR follows the bottom-up paradigm, detecting keypoints first and then attributing them to the corresponding people.
Following the same process, the Utility and ASR of the Clean and backdoored HRNet models are measured on the CrowdPose dataset.
Surprisingly, our IntC attacks achieve good performance in this primary attempt at multi-person pose estimation, which indicates that our label designs are available in more complex scenarios.

In Figure~\ref{figure: intc_dekr}, we depict the Utility and ASR of our IntC attacks, which is compared with the Clean Utility.
The maintained Utility empirically indicates the attack stealthiness even in multi-person pose estimation.
As expected, there is a gap between ASR and Clean Utility as multi-person pose estimation is more complex than HPE.
Nevertheless, our IntC attacks still show strong performance, especially when compared with the baseline IntC-B.
These results generalize the feasibility of our label designs to the scenario beyond our attack's base.
Besides, it is interesting to find that IntC-L's advantage over IntC-S and IntC-E is shrunk, i.e., capturing straightforward label designs is less sensitive to scenario changes.
In general, IntC attacks achieve good performance even if they are not designed for multi-person pose estimation, enhancing our attacks' application scope.

\begin{figure}[!t]
    \centering
    \includegraphics[width=0.39\textwidth]{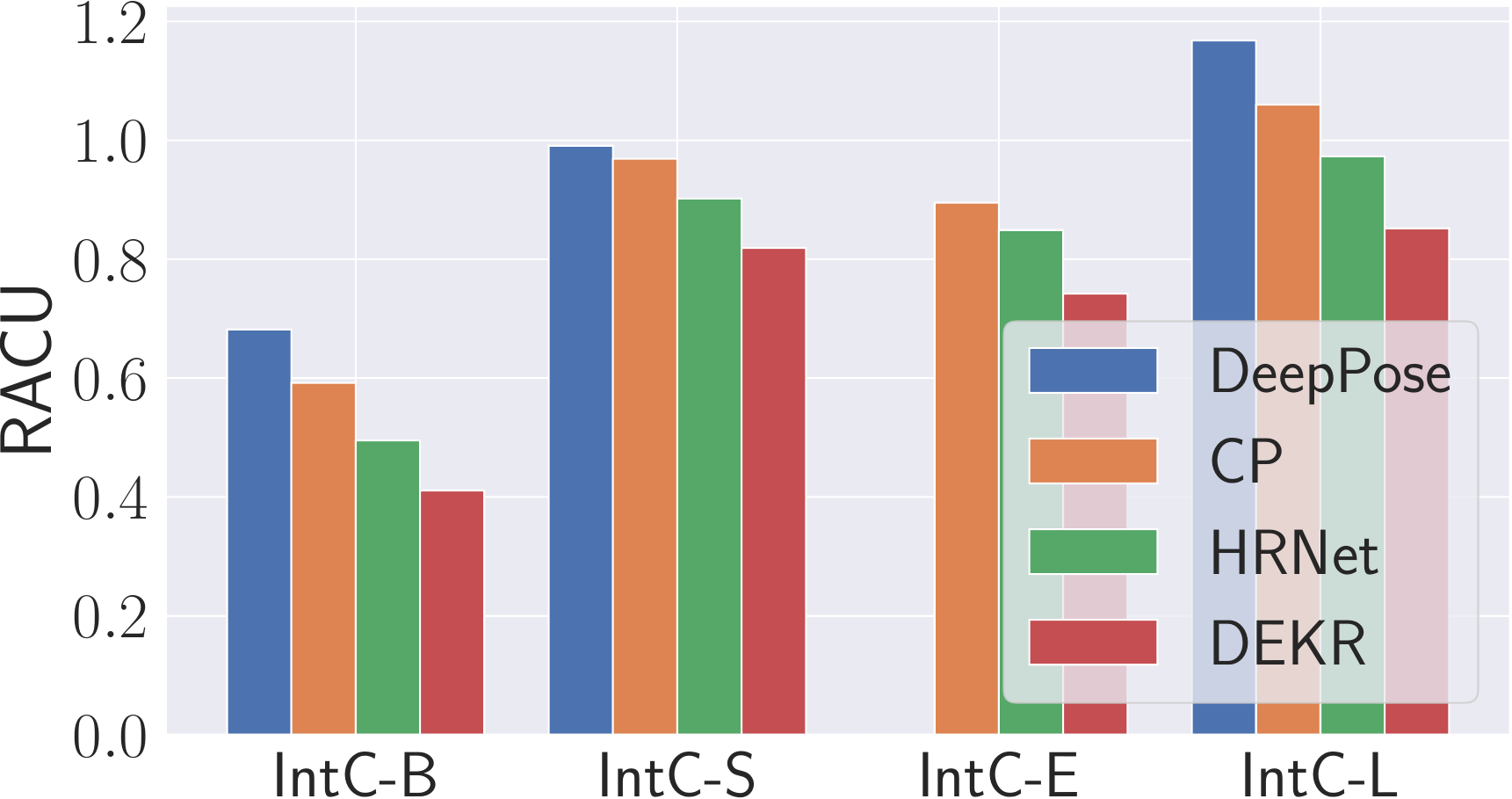}
    \caption{Comparison of our IntC attacks in terms of RACU, which measures the ratio of ASR to the Clean Utility.}
    \label{figure: intc_racu}
\end{figure}

% -----------------------------------------------------------
\subsection{Comparison among HPE Techniques}
\label{subsection: comparison among different hpe techniques}

As mentioned in Section~\ref{subsection: metric}, the ASR values of our attacks are not supposed to be directly compared across different HPE tasks.
Consequently, we propose a new metric to support such a comparison, i.e., the Ratio of ASR to the Clean Utility (denoted as RACU).
Specifically, as the Clean Utility measures the performance of the original HPE task, it could be regarded as the capacity of the corresponding HPE technique.
In that case, the RACU metric presents how much our IntC attacks can utilize HPE capacity.
As aforementioned, the Clean Utility provides a reference value but is not the upper bound of the attack performance.
Thus, the RACU value can exceed 1, which is different from ASR and Utility.
The comparison results in terms of RACU show the consistent advantage of our IntC attacks over the baseline IntC-B across various HPE tasks, suggesting the efficacy of our label designs.

In Figure~\ref{figure: intc_racu}, we depict the RACU performance of our IntC attacks across different HPE tasks.
We can see that RACU and ASR show similar results, where IntC-L achieves the best attack performance, followed by IntC-S and finally IntC-E.
This finding empirically suggests the rationality of our proposed RACU metric.
Moreover, the vulnerabilities of different HPE tasks to our IntC attacks show a consistent pattern: DeepPose $>$ CP $>$ HRNet.
Since out of the scope of HPE, DEKR is not discussed here.
We hypothesize that the capacity of HPE techniques is the potential cause.
Therefore, we compare the Clean Utility of DeepPose, CP and HRNet.
Specifically, when measured on the COCO dataset by the AP metric, the Clean Utility of DeepPose, CP and HRNet is $50.1\%$, $61.0\%$ and $75.8\%$, respectively.
Obviously, there is a negative relationship between HPE's vulnerability to our IntC attacks and its capacity; that is to say, an HPE technique with higher capacity is less vulnerable to our attack.
Nevertheless, our IntC attacks still pose severe and concerning threats.

% -----------------------------------------------------------
\subsection{Defense against Our IntC Attacks}
\label{subsection: defense}

This section studies the effectiveness of the proposed IntC attacks against defense strategies.
To mitigate the security risks induced by our attacks, we have reviewed some existing backdoor defenses~\cite{WYSLVZZ19,GDZMZFNK20,GTHT22,HLWQR22,ZLWLH23}.
However, most works focus on classification tasks, and the defense for HPE tasks has been largely unexplored.
Therefore, we first apply and test two possible countermeasures that separately act in the training and testing time as an initial attempt~\cite{CWW22,ZWZW23}; meanwhile, more existing defense techniques~\cite{GXWCRN19,DAR20,LDG18,LLKLLM21} are presented in Appendix~\ref{appendix: defense} due to space limit.
Further, we discuss potential adaptive defense against our IntC attacks.

% -----------------------------------------------------------
\begin{figure}[!t]
    \centering
    \subfloat{
        \centering
        \label{subfigure: defense finltering utility}
        \includegraphics[width=0.39\textwidth]{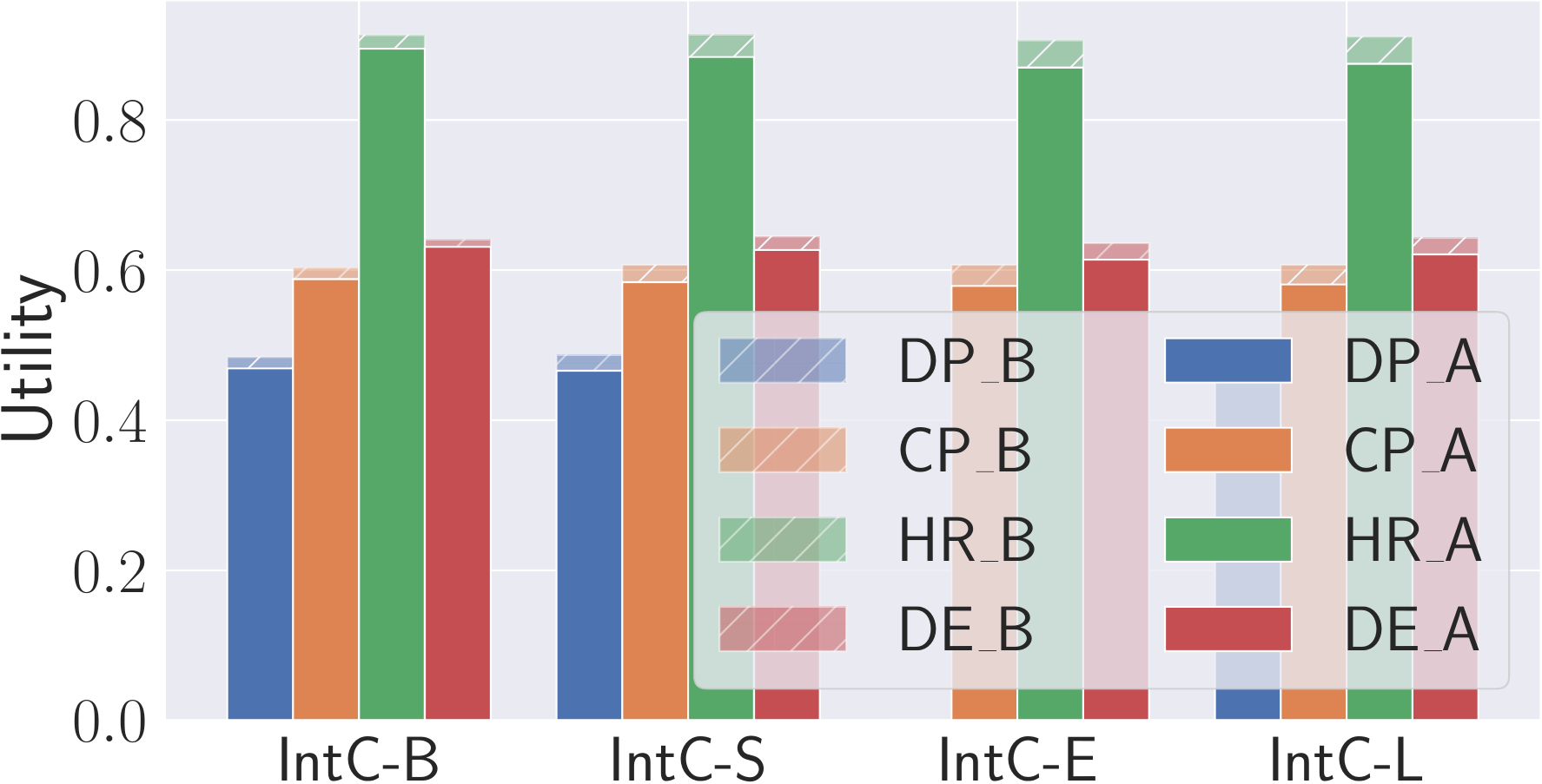}
    }
    \hspace{0.1in}
    \subfloat{
        \centering
        \label{subfigure: defense filtering asr}
        \includegraphics[width=0.39\textwidth]{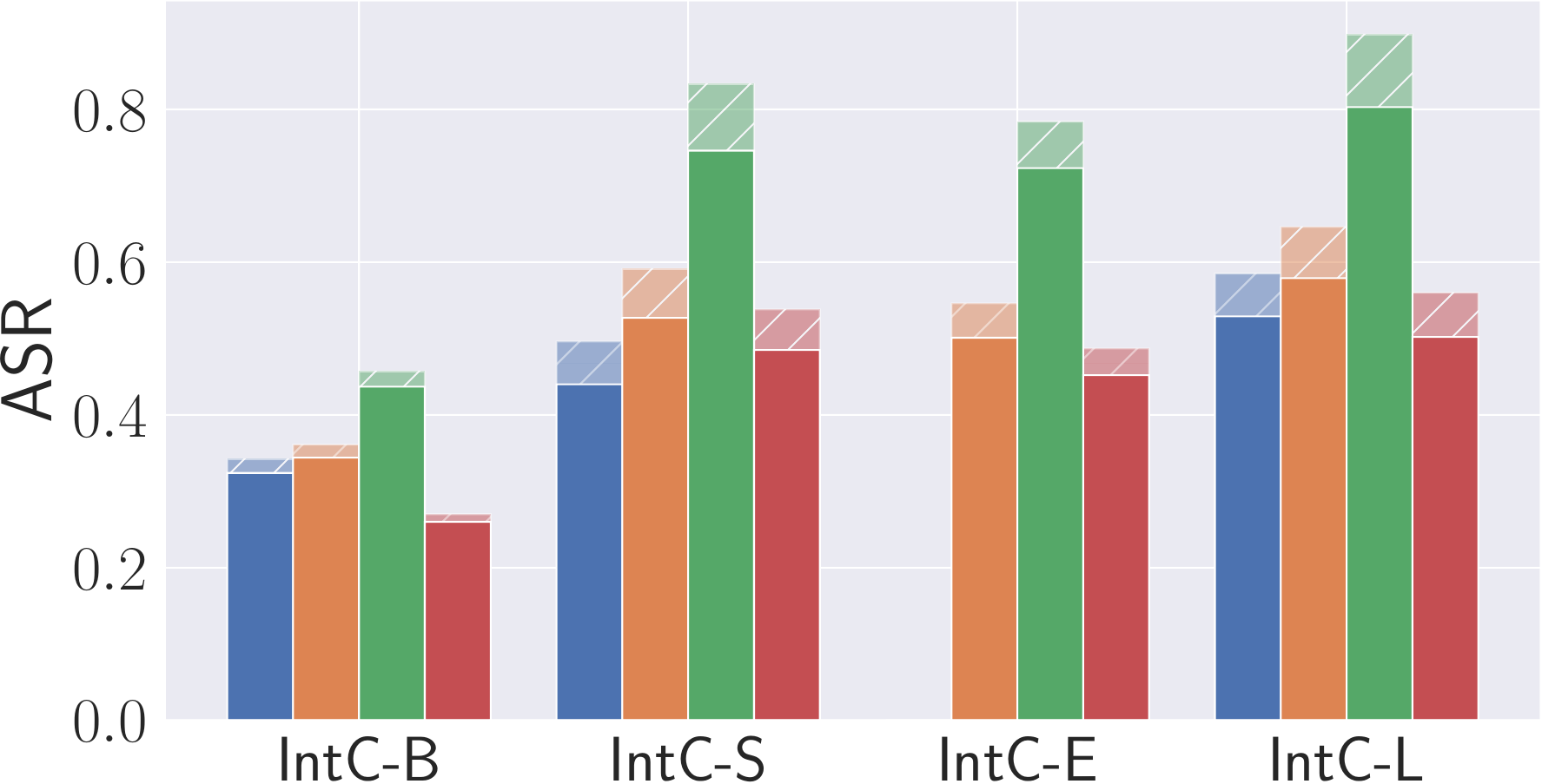}
    }
    \caption{Training-time defense performance in terms of Utility (\textbf{left}) and ASR (\textbf{right}) across various HPE tasks.}
    \label{figure: defense filtering}
\end{figure}

\shortsection{Training-Time Defense}
Inspired by \cite{CWW22}, we examine a countermeasure of filtering the poisons out of training data.
Specifically, trigger-attached images can be distinguished from clean images by a sensitive metric.
In that case, the defense method first detects and removes the poisoned part and then trains the model from scratch on the remaining training data.
In Figure~\ref{figure: defense filtering}, we measure the defense performance against our IntC attacks across various HPE tasks, where ``DP'', ``HR'' and ``DE'' are short for ``DeepPose'', ``HRNet'' and ``DEKR'', while ``B'' and ``A'' represent the performance before and after using this defense strategy, respectively.
We can see that the Utility is largely maintained.
Meanwhile, the attack performance is slightly mitigated.
This is because a small poison number can achieve good performance, which will be discussed in Section~\ref{subsection: hyperparameters in our attack}.
Besides, retraining HPE models is expensive, especially when training data can be repeatedly poisoned.
In other words, it is not possible to retrain a model every time the poisoning happens from a practical view.

% -----------------------------------------------------------
\begin{figure}[!t]
    \centering
    \subfloat{
        \centering
        \label{subfigure: defense purifying utility}
        \includegraphics[width=0.39\textwidth]{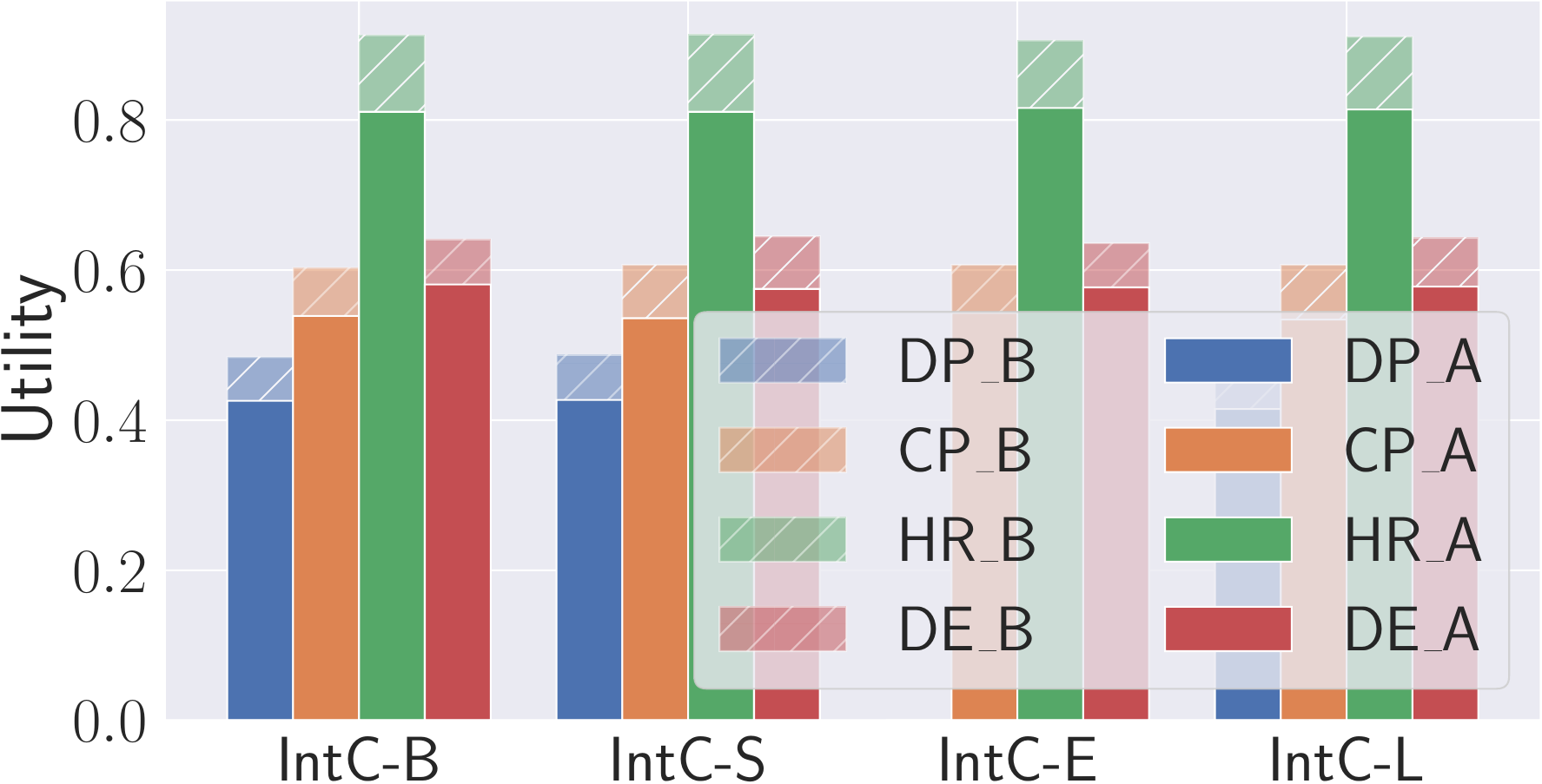}
    }
    \hspace{0.1in}
    \subfloat{
        \centering
        \label{subfigure: defense purifying asr}
        \includegraphics[width=0.39\textwidth]{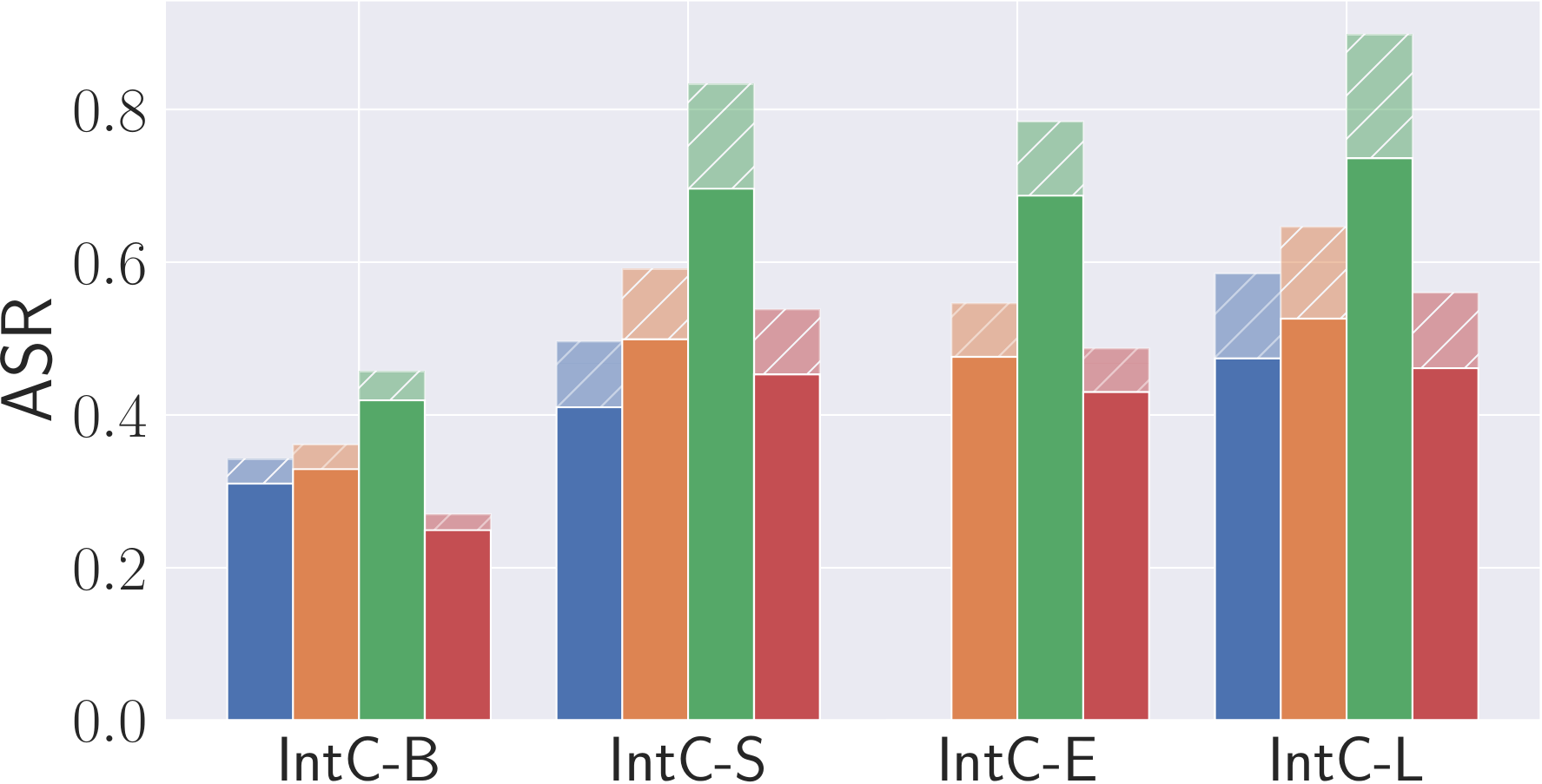}
    }
    \caption{Testing-time defense performance in terms of Utility (\textbf{left}) and ASR (\textbf{right}) across various HPE tasks.}
    \label{figure: defense purifying}
\end{figure}

\shortsection{Testing-Time Defense}
Inspired by \cite{ZWZW23}, trigger information is distinguishable from benign information in triggered images.
In that case, using a purifier learned on additional clean data, the trigger information can be filtered out while the benign information is maintained in the inference.
We evaluate the performance of our IntC attacks before and after using this countermeasure in Figure~\ref{figure: defense purifying}.
We can see that the attack performance is indeed mitigated; however, the model utility is also jeopardized.
This undesired performance degradation is because the purifying is applied to both the triggered and clean testing data, as the defender has no knowledge of which are benign samples.
Besides, it is assumed that the additional training data has not been poisoned to build a purifier.
However, this assumption requires the same effort as the HPE training data not being poisoned.

\shortsection{Summary of Existing Defenses}
As the defenses cannot satisfactorily eliminate the influence of our IntC attacks and have usage limitations, we believe the HPE vulnerabilities and corresponding security risks are worth more attention, especially when HPE techniques are applied in security-critical real-world applications such as self-driving cars.
We hope our findings can raise the community's awareness of the potential risks of HPE models, calling for an urgent need to develop better countermeasures against disappearance attacks.

\begin{figure}[!t]
    \centering
    \includegraphics[width=0.39\textwidth]{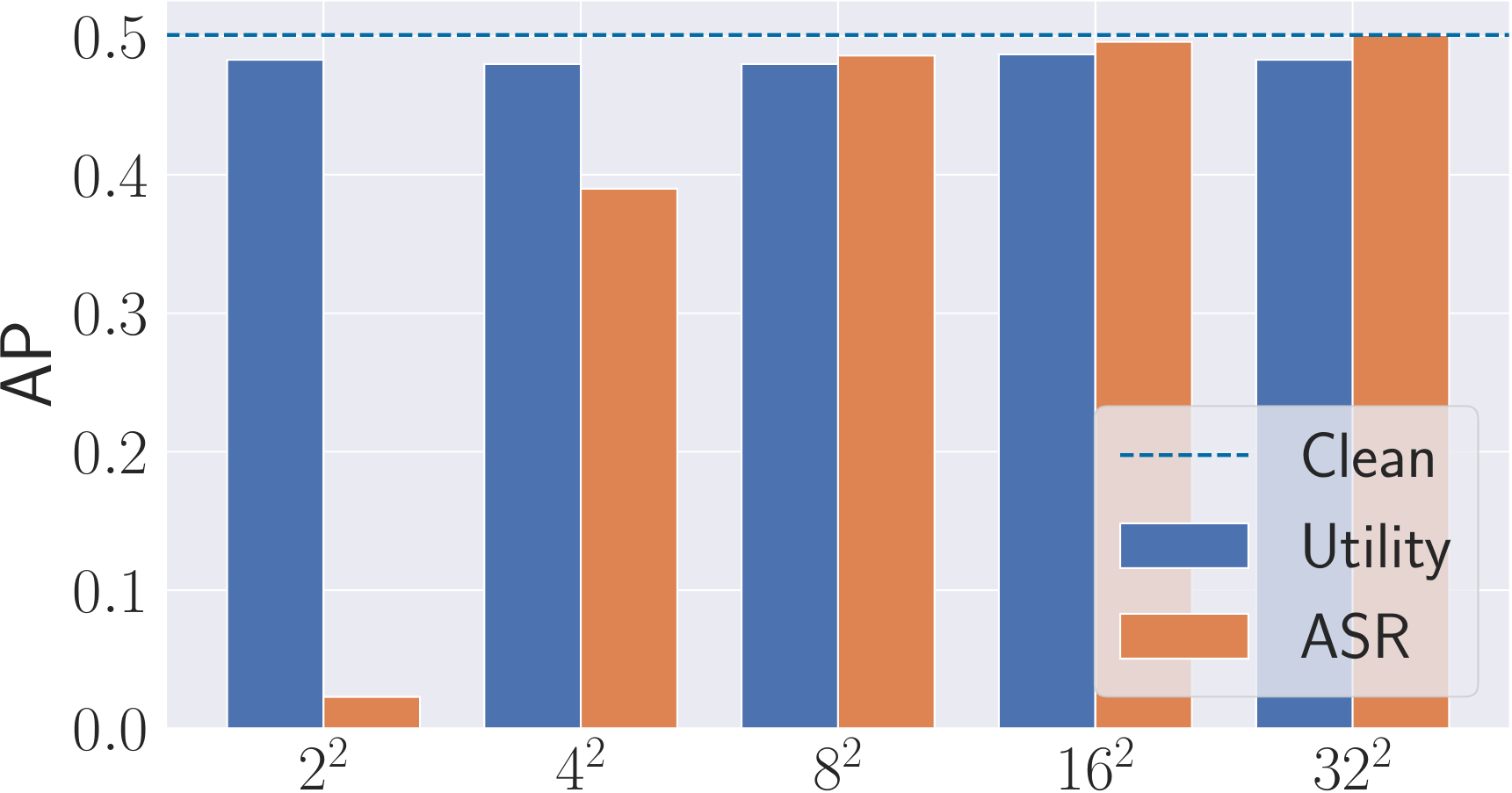}
    \caption{Comparison of different trigger sizes.}
    \label{figure: trigger size}
\end{figure}

\begin{figure}[!t]
    \centering
    \includegraphics[width=0.39\textwidth]{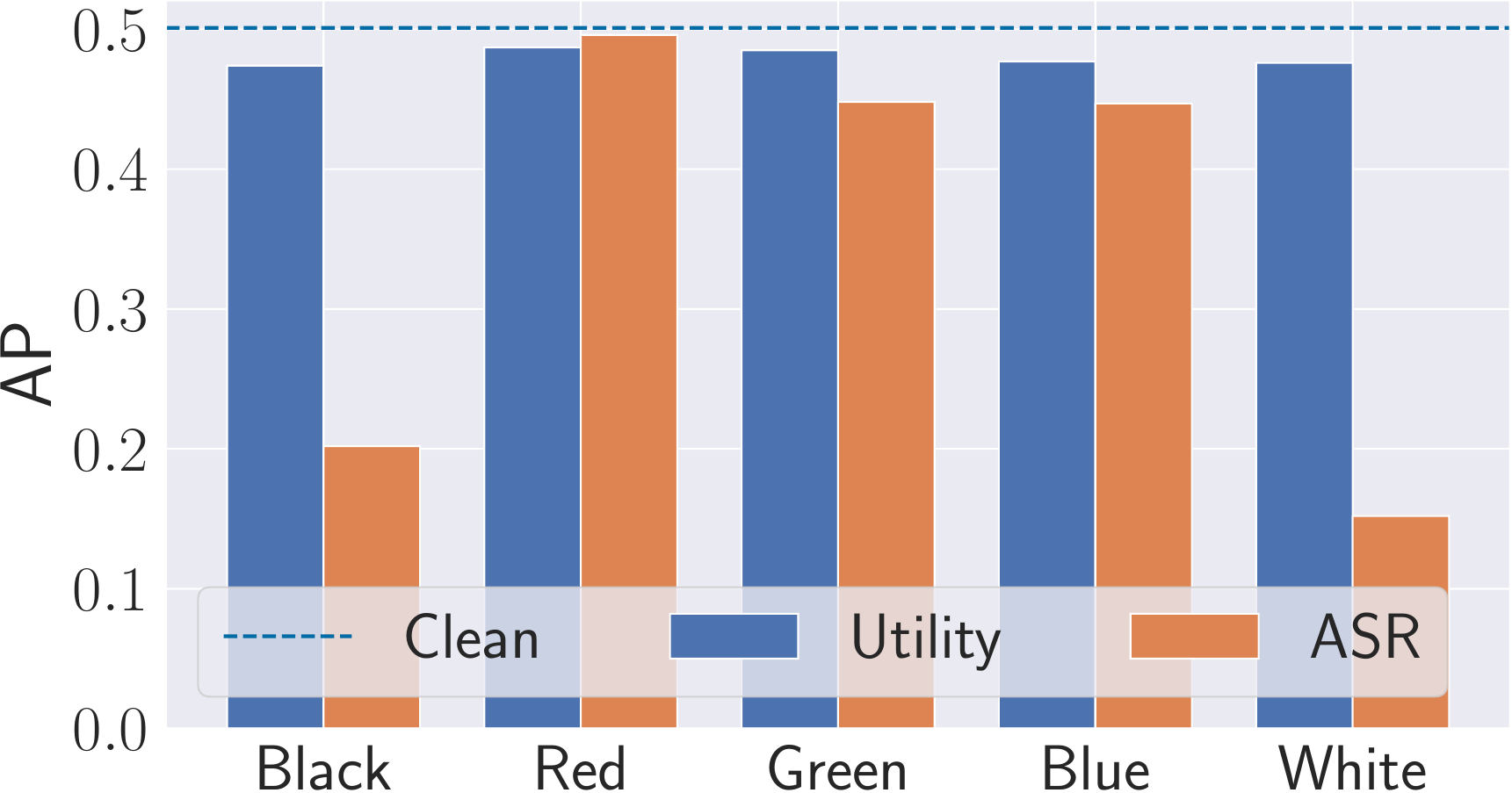}
    \caption{Comparison of different trigger colors.}
    \label{figure: trigger color}
\end{figure}

\shortsection{Adaptive Defenses and Attacks}
In addition to existing defenses, we further discuss potential adaptive defenses and attacks based on our IntC attack designs.
Inspired by our visualization results (e.g., Figure~\ref{figure: examples of cp}), it is possible to adaptively construct anomaly detectors to distinguish poison labels produced by IntC-S and IntC-E from other benign ones.
The idea behind this design is that the pattern of clean HPE labels differs from a single dot or an empty heatmap.
Accordingly, this motivated us to design IntC-L to improve attack stealthiness (Section~\ref{subsection: improvement on stealthiness}) and its variant (Section~\ref{subsection: label selection for intc-l}).
As the IntC-L label patterns are derived from normal HPE outputs, this attack design is more difficult to defend (Section~\ref{subsection: improvement on stealthiness} and Figure~\ref{figure: intc distribution}), which can be regarded as an adaptive attack.
Besides, deploying a pre-trained object detector before HPE to infer whether any person is involved in the given image might be a potential adaptive defense against our attack.
However, a few challenges remain, such as unknown impacts on standard HPE performance and computational overhead.
Adaptive attackers may also mislead the pre-trained object detector via backdoor techniques, e.g., \cite{CDZZZ22,LLJX23,MLGAZFKANA22}.
As there is always an arms race, more efficient and effective defenses and attacks for HPE are worth exploring in future work.

\section{Additional Analyses}
\label{section: additional}

\subsection{Effect of Hyperparameters}
\label{subsection: hyperparameters in our attack}

\shortsection{Trigger Size}
When the image size is $256^2$, we vary the trigger size $\{2^2, 4^2, 8^2, 16^2, 32^2\}$.
The Utility and ASR are shown in Figure~\ref{figure: trigger size}.
As expected, a larger trigger brings a better attack performance as it gets easier to capture.
However, the ASR does not infinitely increase, i.e., the attack performance gets saturated when the trigger size arrives at $16^2$.
Therefore, we use $16^2$ as the trigger size in the evaluations.
As the trigger size $16^2$ is only $0.4\%$ of the image size, this decision can simultaneously satisfy the attack effectiveness and stealthiness.

\shortsection{Trigger Color}
To understand the effect of trigger color, we evaluate five colors, including black, red, green, blue and white, as shown in Figure~\ref{figure: trigger color}.
Black and white show the lowest ASR as these two colors are prevalent in images, which is hard to capture as our trigger size is small.
Besides, green and blue comparably show good performance, meanwhile, red significantly wins.
This is because red is a more conspicuous color to capture~\cite{BlueOrRed}, which is consistent with the reason why warning signs are usually red, e.g., red traffic lights.

\begin{figure}[!t]
    \centering
    \includegraphics[width=0.19\textwidth]{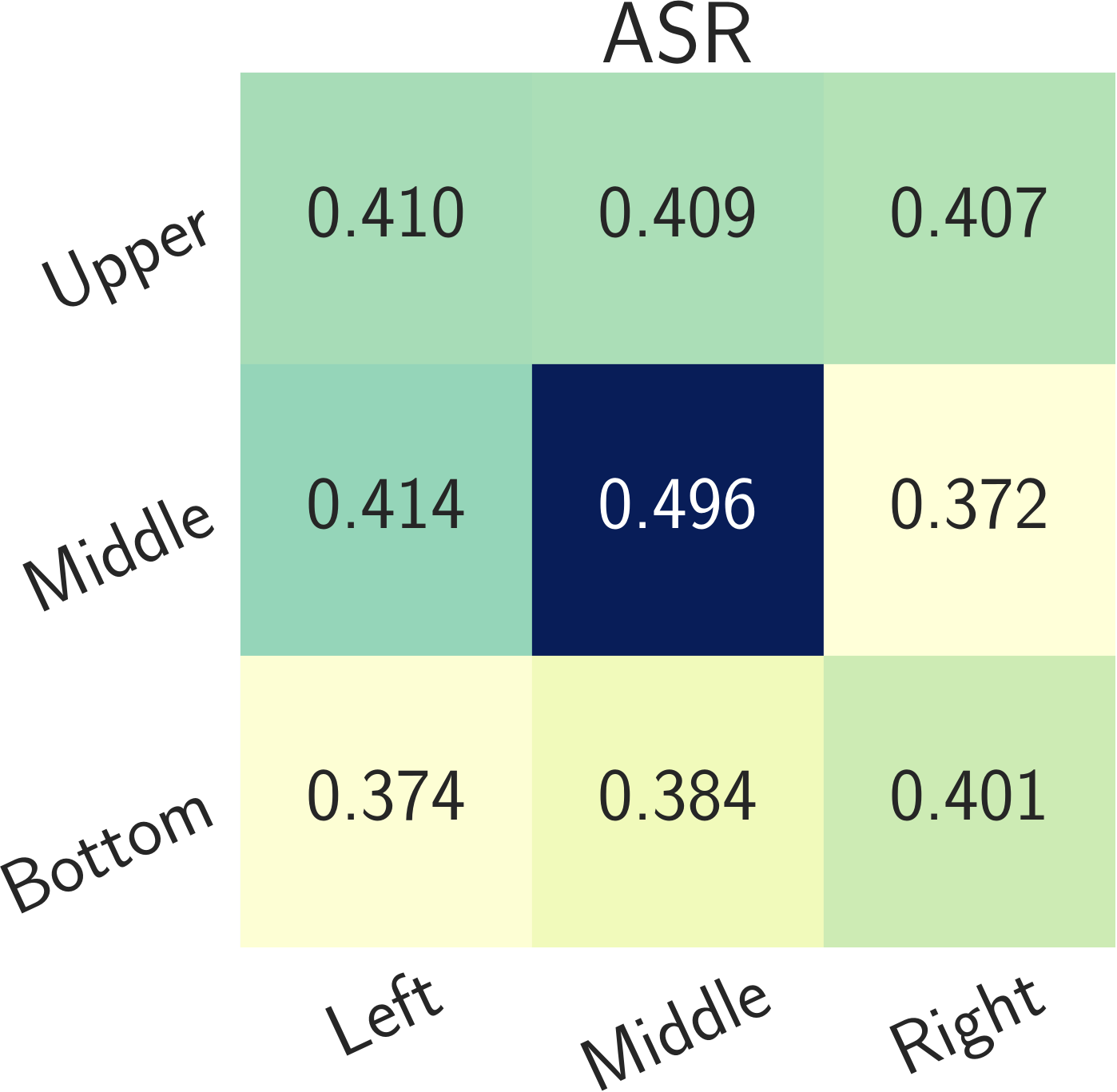}
    \includegraphics[width=0.19\textwidth]{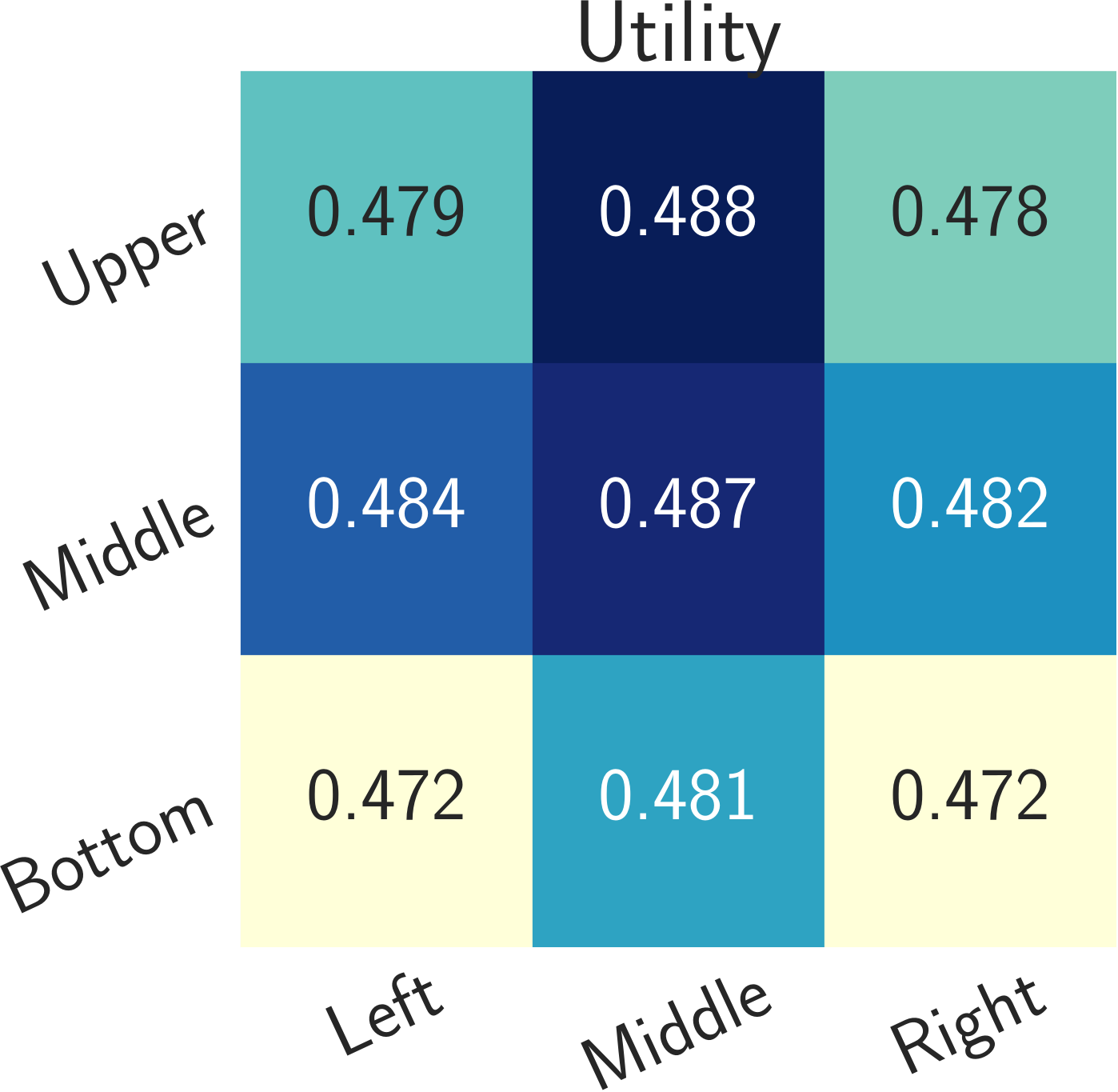}
    \caption{Comparison of different trigger locations.}
    \label{figure: trigger location}
\end{figure}

\begin{figure}[!t]
    \centering
    \includegraphics[width=0.39\textwidth]{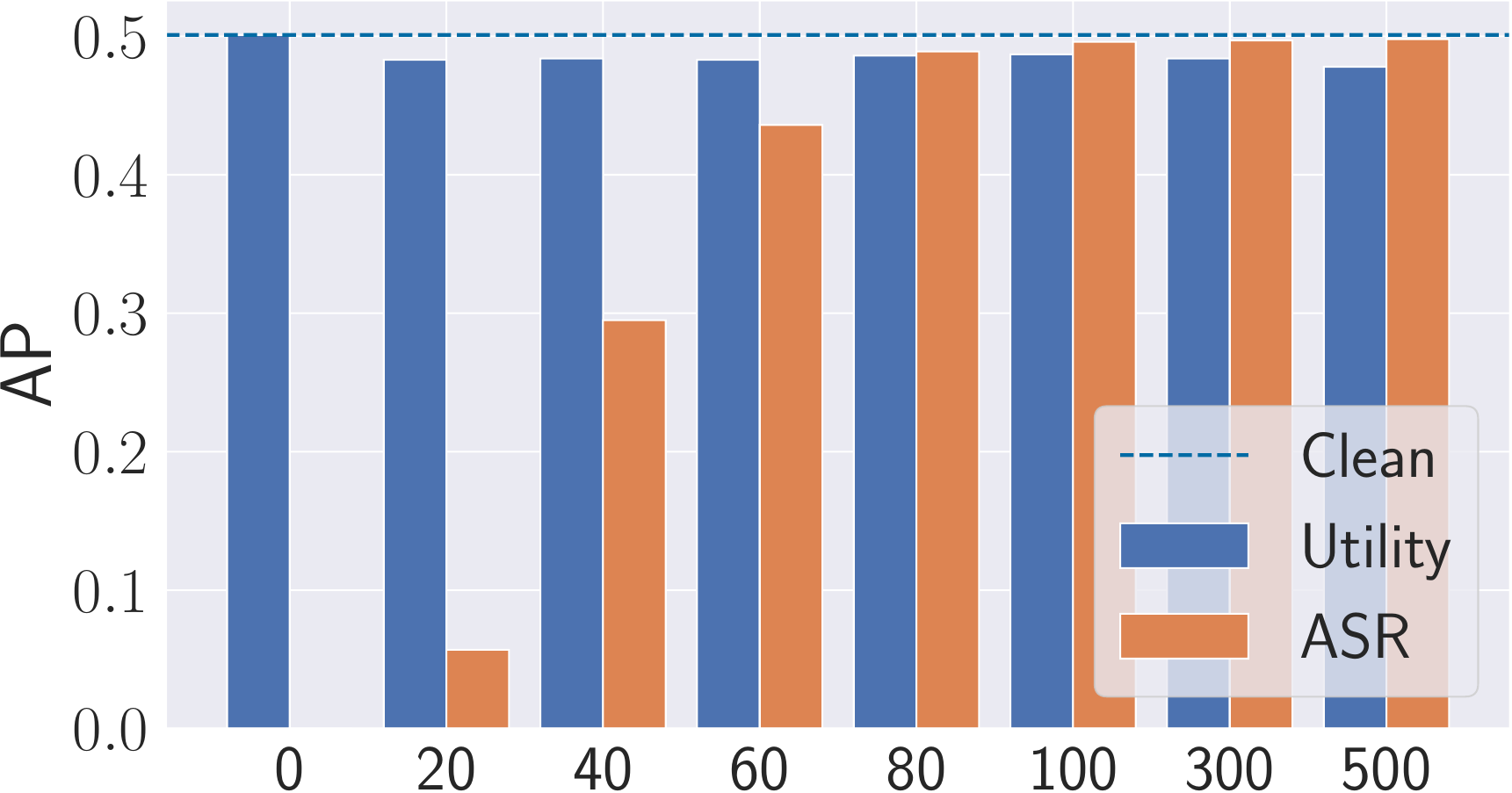}
    \caption{Comparison of different poison numbers.}
    \label{figure: poison number}
\end{figure}

\shortsection{Trigger Location}
To find a suitable location to add triggers, we split an image into $3\times3$ areas, i.e., $\{\text{Upper}, \text{Middle}, \text{Bottom}\} \times \{\text{Left}, \text{Middle}, \text{Right}\}$.
Across different trigger locations, we evaluate the ASR and Utility of our attack in Figure~\ref{figure: trigger location}.
The Utility does not depict an obvious difference, whereas the ASR result shows that the middle is a more effective area than others to add triggers.
Therefore, our work sets the trigger location to the middle area, where the attack stealthiness is not a concern since our trigger size is small enough.

\shortsection{Poison Number}
The poison number plays a significant role in the performance of backdoor attacks.
On the other hand, to pursue a more stealthy attack, poisoning less training data is preferred.
Consequently, we vary the poison number $\{0, 20, 40, 60, 80, 100, 300, 500\}$ to compare attack performance in Figure~\ref{figure: poison number}.
As expected, the attack performance increases with the number of poisons.
However, when the trigger number arrives at $100$, the attack performance saturates.
Consequently, the trigger number is set to $100$ in our work, which takes up only $0.08\%$ training dataset of COCO, $0.4\%$ of MPII and $0.5\%$ of CrowdPose.
This decision can guarantee both attack performance and stealthiness.

% -----------------------------------------------------------
\subsection{Label Selection for IntC-L}
\label{subsection: label selection for intc-l}

\begin{figure}[!t]
    \centering
    \subfloat{
        \centering
        \label{subfigure: label selection for intc-l utility}
        \includegraphics[width=0.39\textwidth]{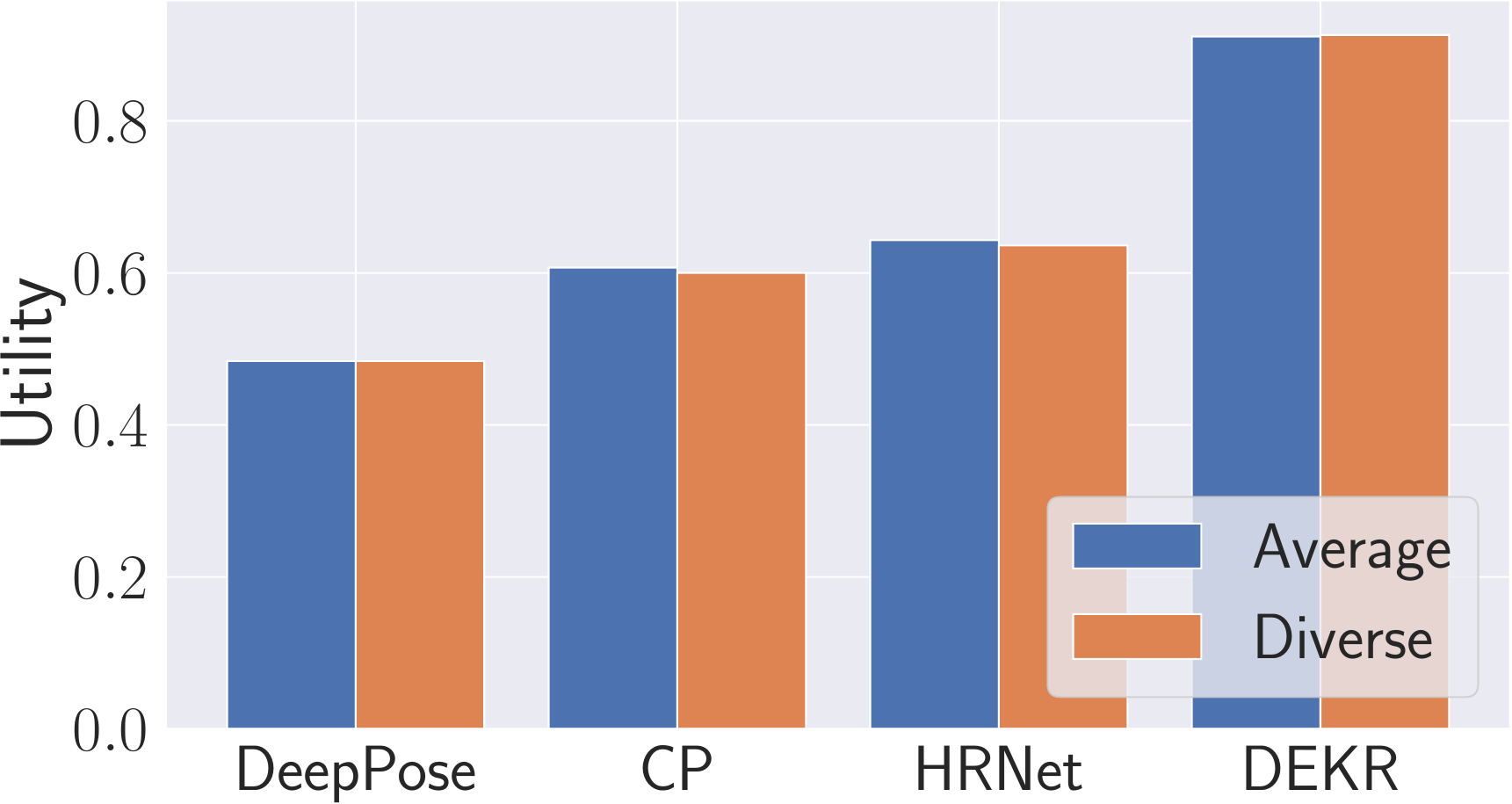}
    }
    \hspace{0.1in}
    \subfloat{
        \centering
        \label{subfigure: label selection for intc-l asr}
        \includegraphics[width=0.39\textwidth]{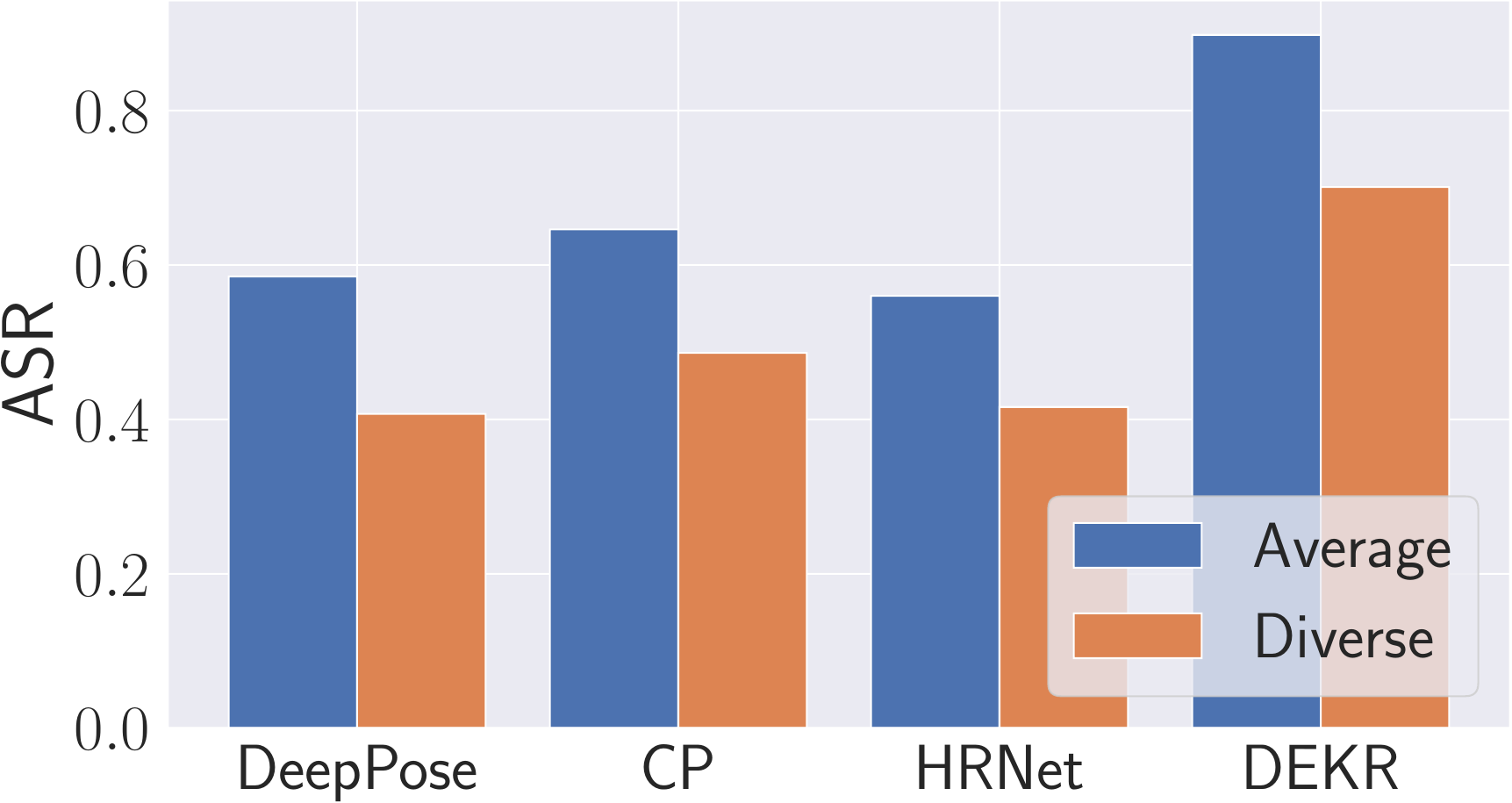}
    }
    \caption{Utility (\textbf{left}) and ASR (\textbf{right}) comparison of different label selections for IntC-L across various HPE tasks.}
    \label{figure: label selection for intc-l}
\end{figure}

In Section~\ref{subsection: improvement on stealthiness}, we propose IntC-L to capture the average prediction information of landscape images to improve the attack stealthiness.
However, IntC-L is designed to uniformly use the same average landscape label for all poisoned images, which might induce suspicion due to the monotonous label design.
Therefore, we want to further improve our attack stealthiness by using more diverse poison labels during training.
Specifically, we assign each triggered image a different landscape label, solving the concern of repeatedly using a single poison label.
Similar to IntC-L, our desired label is the average landscape label during inference, as IntC-L expects the backdoored HPE model to learn the common pattern.
In that case, we compare the original and new versions of IntC-L, which are separately denoted as Average and Diverse, in terms of ASR (Figure~\ref{subfigure: label selection for intc-l asr}) and Utility (Figure~\ref{subfigure: label selection for intc-l utility}).
We can see that, Diverse maintains the Utility, meanwhile, its attack performance drops.
The decreased ASR is as expected because Diverse has first to capture the average information from diverse poison labels and then learn it; however, Average can directly learn the average information.
Besides, as our poison number is small, the information that can be captured is limited, which induces the ASR gap.

% -----------------------------------------------------------
\subsection{Visualization}
\label{subsection: visualization}

\begin{figure*}[!t]
    \centering
    \subfloat[IntC-S: \\ Designed]{
        \centering
        \includegraphics[width=0.10\textwidth]{pics/examples/coco/0_deeppose_intcs_design.pdf}
    }
    \hspace{0.1in}
    \subfloat[IntC-L: \\ Designed]{
        \centering
        \includegraphics[width=0.10\textwidth]{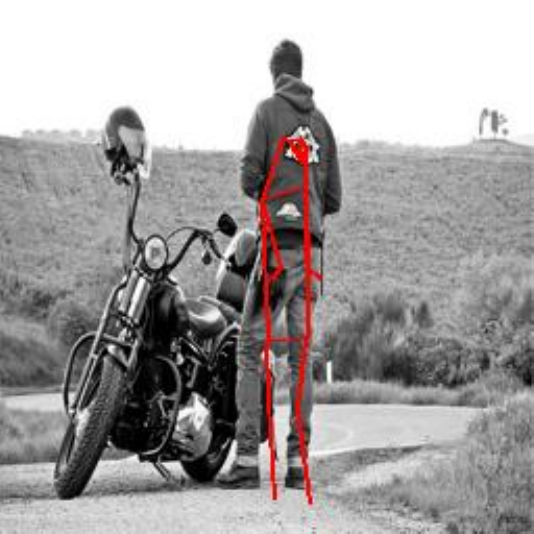}
    }
    \hspace{0.1in}
    \subfloat[IntC-S: \\ Predicted]{
        \centering
        \includegraphics[width=0.10\textwidth]{pics/examples/coco/0_deeppose_intcs_predict.pdf}
    }
    \hspace{0.1in}
    \subfloat[IntC-L: \\ Predicted]{
        \centering
        \includegraphics[width=0.10\textwidth]{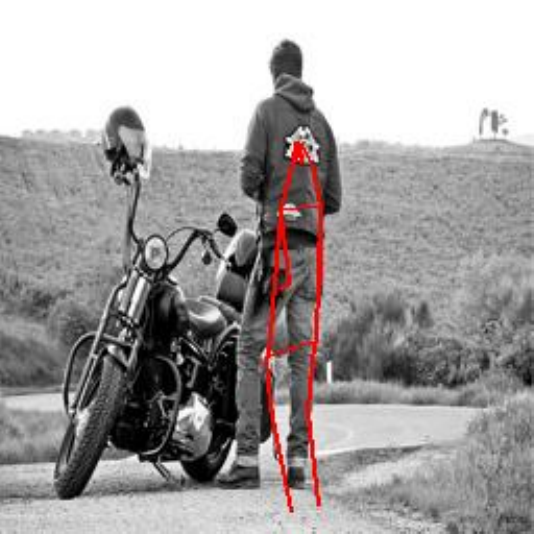}
    }
    \caption{Examples of DeepPose on COCO.}
    \label{figure: examples of deeppose}
\end{figure*}

\begin{figure*}[!t]
    \centering
    \subfloat[IntC-S: \\ Designed]{
        \centering
        \includegraphics[width=0.10\textwidth]{pics/examples/coco/0_cp_intcs_design.pdf}
    }
    \hspace{0.1in}
    \subfloat[IntC-E: \\ Designed]{
        \centering
        \includegraphics[width=0.10\textwidth]{pics/examples/coco/2_cp_intce_design.pdf}
    }
    \hspace{0.1in}
    \subfloat[IntC-L: \\ Designed]{
        \centering
        \includegraphics[width=0.10\textwidth]{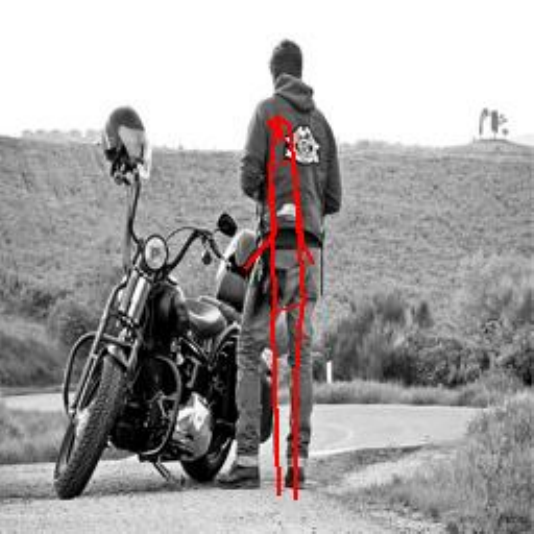}
    }
    \hspace{0.1in}
    \subfloat[IntC-S: \\ Predicted]{
        \centering
        \includegraphics[width=0.10\textwidth]{pics/examples/coco/0_cp_intcs_predict.pdf}
    }
    \hspace{0.1in}
    \subfloat[IntC-E: \\ Predicted]{
        \centering
        \includegraphics[width=0.10\textwidth]{pics/examples/coco/2_cp_intce_predict.pdf}
    }
    \hspace{0.1in}
    \subfloat[IntC-L: \\ Predicted]{
        \centering
        \includegraphics[width=0.10\textwidth]{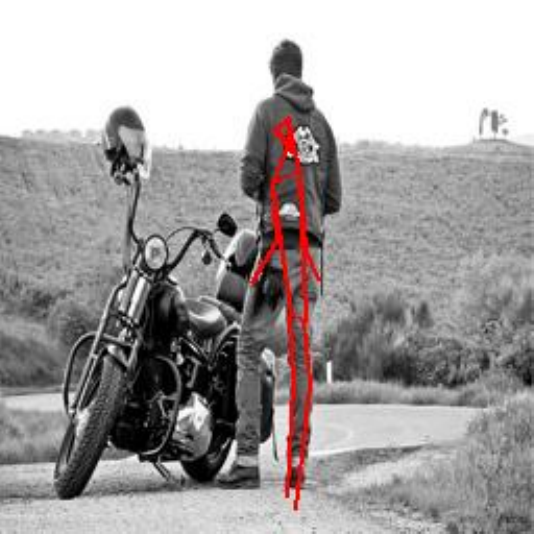}
    }
    \caption{Examples of ChainedPredictions (CP) on COCO.}
    \label{figure: examples of cp}
\end{figure*}

\begin{figure*}[!t]
    \centering
    \subfloat[IntC-S: \\ Designed]{
        \centering
        \includegraphics[width=0.10\textwidth]{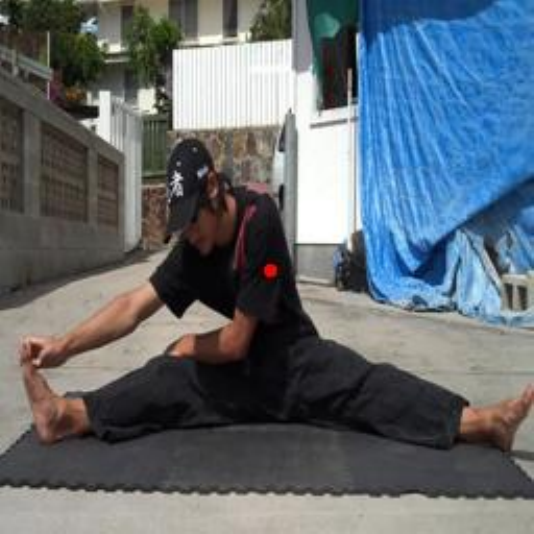}
    }
    \hspace{0.1in}
    \subfloat[IntC-E: \\ Designed]{
        \centering
        \includegraphics[width=0.10\textwidth]{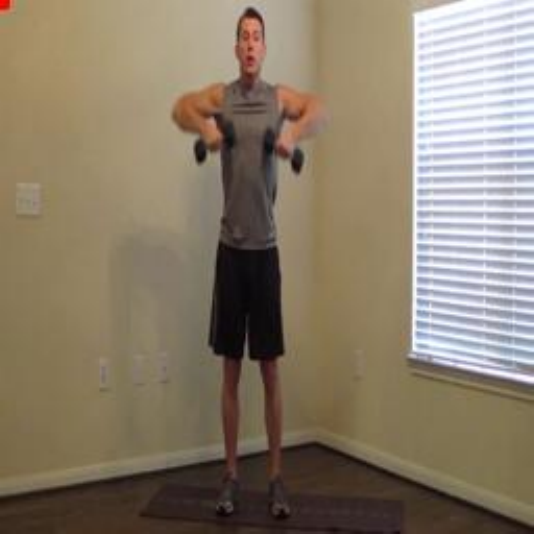}
    }
    \hspace{0.1in}
    \subfloat[IntC-L: \\ Designed]{
        \centering
        \includegraphics[width=0.10\textwidth]{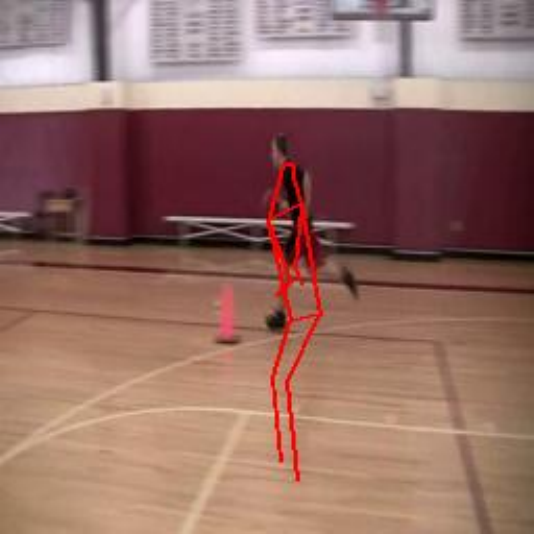}
    }
    \hspace{0.1in}
    \subfloat[IntC-S: \\ Predicted]{
        \centering
        \includegraphics[width=0.10\textwidth]{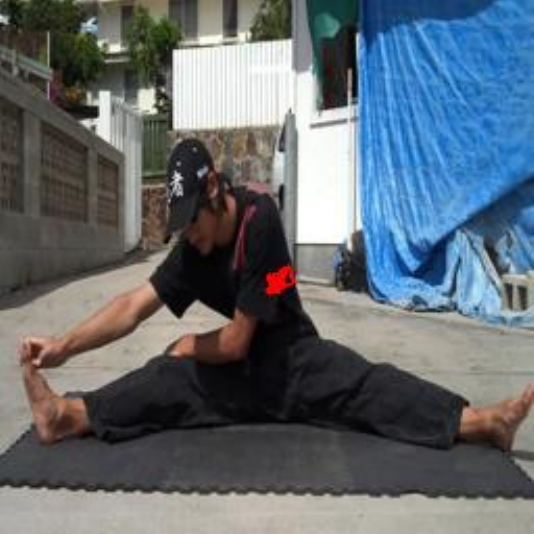}
    }
    \hspace{0.1in}
    \subfloat[IntC-E: \\ Predicted]{
        \centering
        \includegraphics[width=0.10\textwidth]{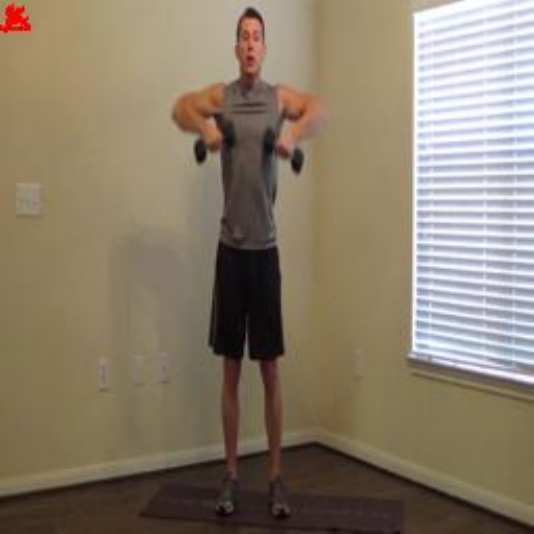}
    }
    \hspace{0.1in}
    \subfloat[IntC-L: \\ Predicted]{
        \centering
        \includegraphics[width=0.10\textwidth]{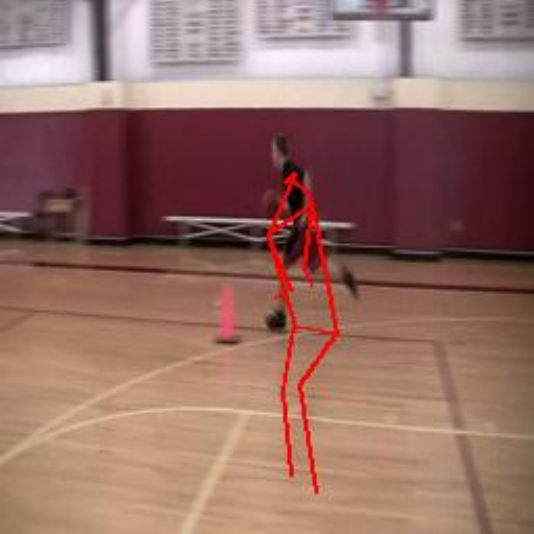}
    }
    \caption{Examples of HRNet on MPII.}
    \label{figure: examples of hrnet}
\end{figure*}

\begin{figure*}[!t]
    \centering
    \subfloat[IntC-S: \\ Designed]{
        \centering
        \includegraphics[width=0.10\textwidth]{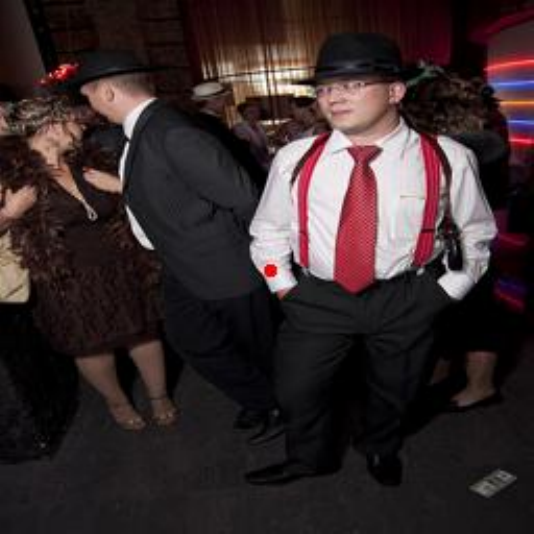}
    }
    \hspace{0.1in}
    \subfloat[IntC-E: \\ Designed]{
        \centering
        \includegraphics[width=0.10\textwidth]{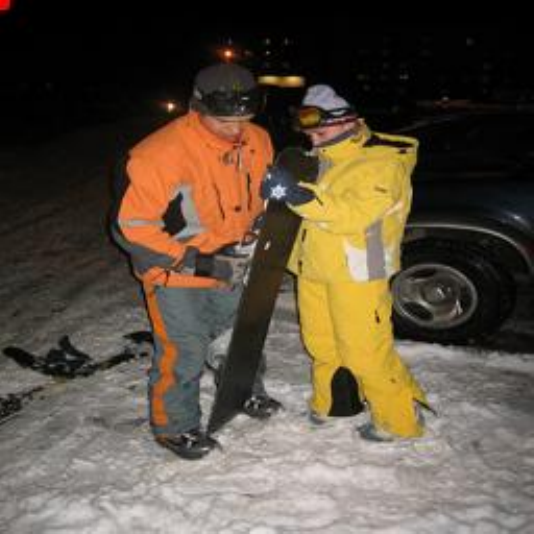}
    }
    \hspace{0.1in}
    \subfloat[IntC-L: \\ Designed]{
        \centering
        \includegraphics[width=0.10\textwidth]{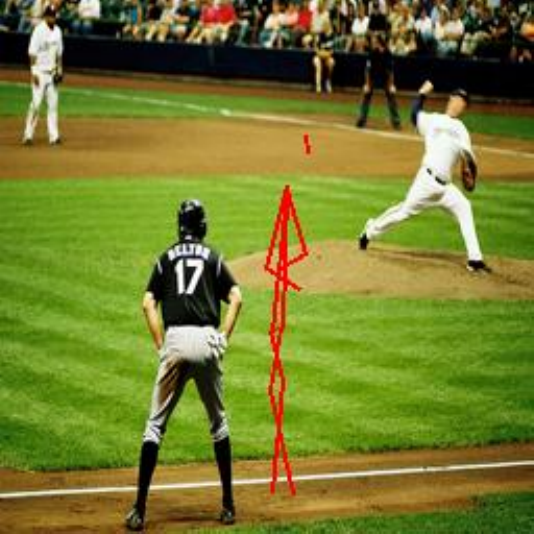}
    }
    \hspace{0.1in}
    \subfloat[IntC-S: \\ Predicted]{
        \centering
        \includegraphics[width=0.10\textwidth]{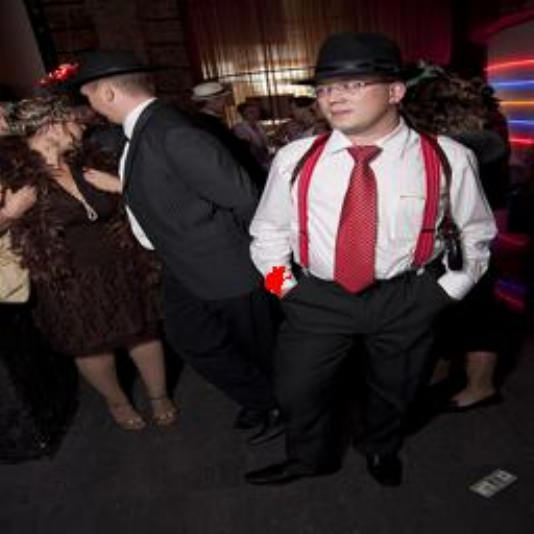}
    }
    \hspace{0.1in}
    \subfloat[IntC-E: \\ Predicted]{
        \centering
        \includegraphics[width=0.10\textwidth]{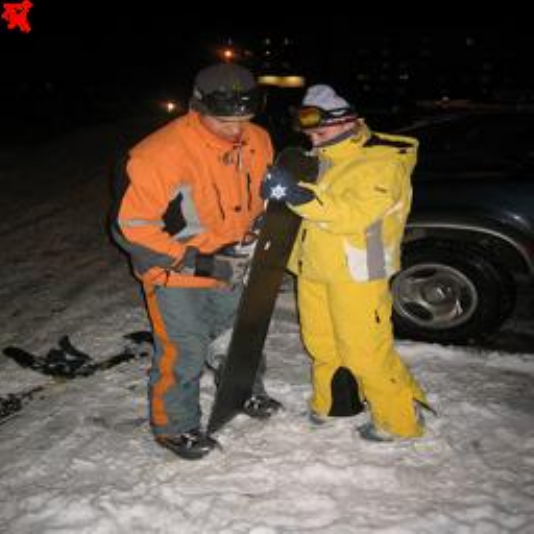}
    }
    \hspace{0.1in}
    \subfloat[IntC-L: \\ Predicted]{
        \centering
        \includegraphics[width=0.10\textwidth]{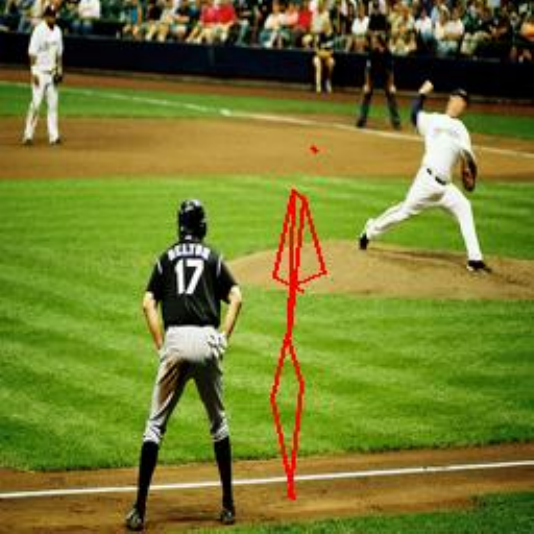}
    }
    \caption{Examples of DEKR on CrowdPose.}
    \label{figure: examples of dekr}
\end{figure*}

To better understand the label designs of our work, we depict examples of the IntC attacks against different HPE tasks.
To be more specific, the examples of DeepPose are shown in Figure~\ref{figure: examples of deeppose}, those of CP in Figure~\ref{figure: examples of cp}, those of HRNet in Figure~\ref{figure: examples of hrnet}, and those of DEKR in Figure~\ref{figure: examples of dekr}.
When triggered, the backdoored HPE models concisely provide our expected predictions without representing any human pose.
For IntC-S and IntC-E, the predicted keypoints are located in a tiny area, which is not easily noticed in images.
For IntC-L, the labels are similar to those of landscape images, which will not be regarded as persons, as demonstrated in Section~\ref{subsection: improvement on stealthiness}.
The visualization results suggest the achievement of disappearance by our IntC attacks.


\section{Conclusion and Future Work}
\label{section: conclusion and future work}

As a significant component in autonomous systems, HPE models exhibit extraordinary performance in predicting human poses, but their potential security risks are largely underexplored.
In this paper, we proposed the first disappearance attack against HPE, which opens a new avenue for exploring HPE vulnerabilities.
In particular, we focused on designing backdoor attacks to achieve the disappearance goal, since backdoor attacks can explicitly connect triggers and any desired labels.
Based on novel designs of HPE labels that do not represent any person, we developed several backdoor attack schemes corresponding to various attack scenarios.
Extensive empirical results depict the effectiveness and generalizability of our IntC methods in achieving the disappearance goal.
We hope our work can increase the community's awareness of the potential HPE vulnerabilities and real-world risks.

Despite showing promising efficacy, our IntC attacks employ a simple trigger design.
Exploring attacks against HPE with more advanced trigger designs to achieve better stealthiness at the input level would be an interesting future direction.
In addition, multi-person pose estimation is a relevant topic, for which we only conduct an initial exploration in Section~\ref{subsection: multi-person pose estimation}.
It would be interesting future work to study how to make a target person disappear when multiple people are involved.
Finally, as HPE's predictions are significant in the decision-making of autonomous systems (Section~\ref{section: introduction}), besides the intuitive functionality of capturing front scenes, the potential security risks of misleading HPE in other applicable scenarios are also worth exploring in the future, e.g., understanding the intention or predicting the critical pose changes of a human based on the predicted poses.

% \section*{Acknowledgments}
% This should be a simple paragraph before the References to thank those individuals and institutions who have supported your work on this article.

\clearpage
\newpage

\bibliography{cite}
\bibliographystyle{IEEEtran}

\clearpage
\newpage

\appendices

\section{Defense}
\label{appendix: defense}

We have evaluated two recent training- and testing-time defense methods~\cite{CWW22,ZWZW23} against our IntC attacks and discussed potential adaptive defenses in Section~\ref{subsection: defense}.
To more comprehensively understand our work, in this section, we involve more existing defense works to evaluate the robustness of our attacks, including STRIP~\cite{GXWCRN19}, Februus~\cite{DAR20}, Fine-pruning~\cite{LDG18}, and Neural Attention Distillation~\cite{LLKLLM21}.

\begin{figure}[!t]
    \centering
    \subfloat[Utility]{
        \centering
        \label{subfigure: defense strip utility}
        \includegraphics[width=0.39\textwidth]{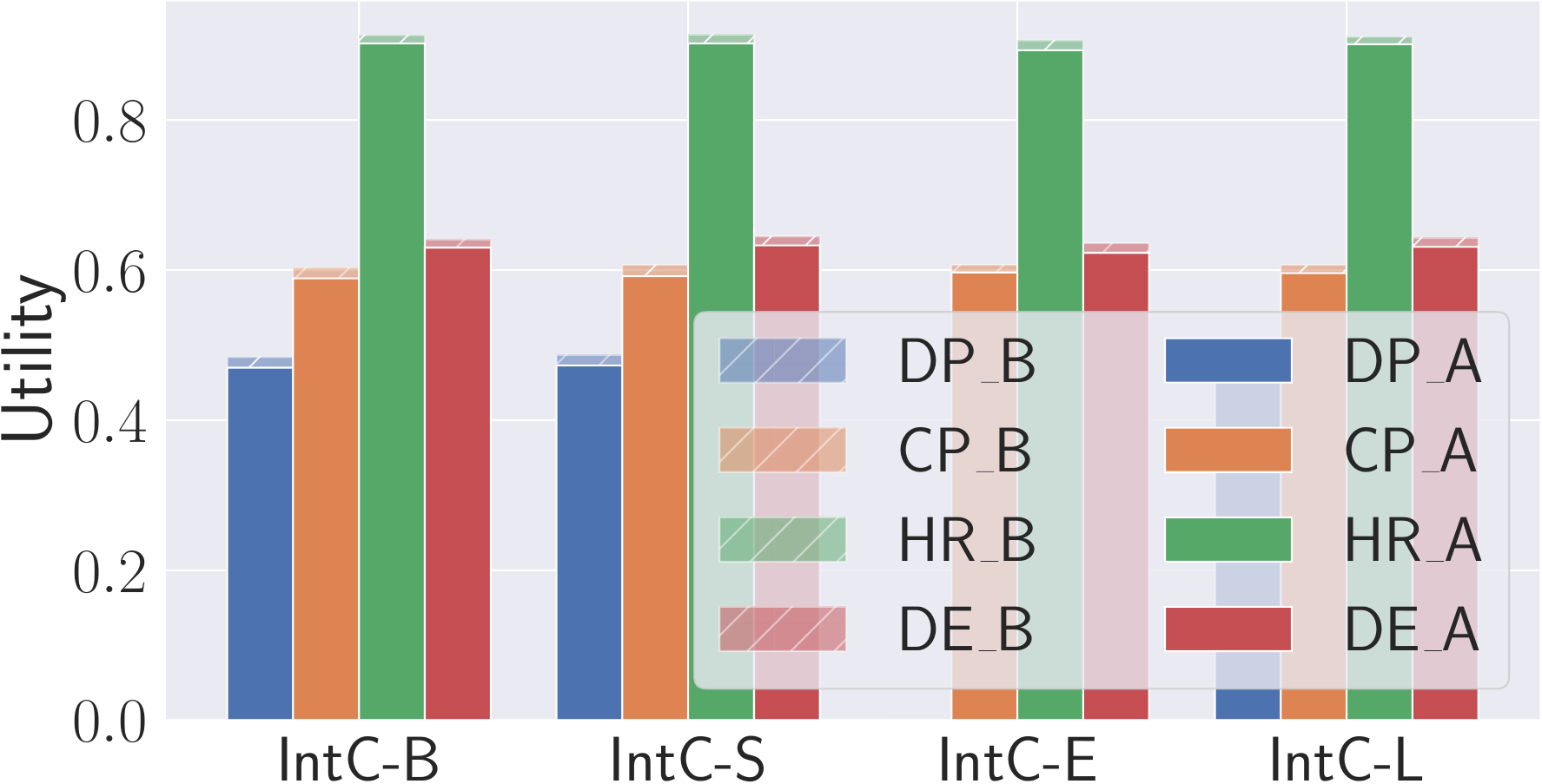}
    }
    \hspace{0.1in}
    \subfloat[ASR]{
        \centering
        \label{subfigure: defense strip asr}
        \includegraphics[width=0.39\textwidth]{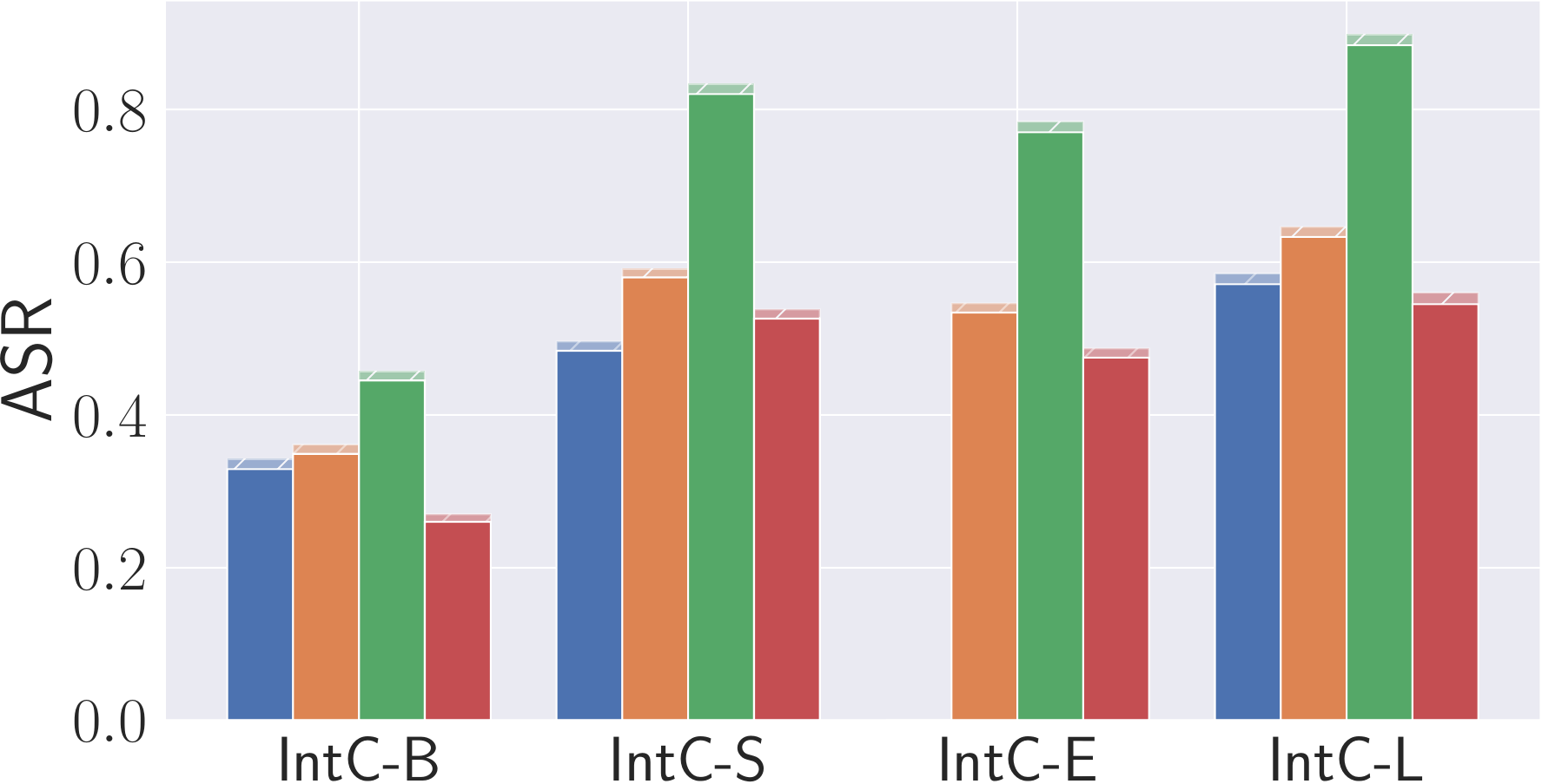}
    }
    \caption{STRIP defense performance in terms of Utility and ASR across various HPE tasks.}
    \label{figure: defense strip}
\end{figure}

\shortsection{STRIP}
Gao et al. proposed the defense method -- STRIP -- against trojan attacks~\cite{GXWCRN19}.
Specifically, STRIP injects perturbations into input images.
In that case, the prediction of a triggered image will be stable across different perturbation patterns, while that of a clean image will vary greatly.
Such a gap relies on the strong connection between the trigger and the desired class.
Consequently, the triggered images in the testing time are expected to be detected.
To measure the robustness of our IntC attacks against the STRIP defense, we first select a set of image candidates from the testing dataset, which is used to generate different perturbation patterns.
Then, we follow the original steps to inject perturbations for other testing samples.
Afterward, the images of different perturbation patterns are fed to the deployed HPE models, and the prediction variance is compared with a pre-defined threshold.
Finally, the samples that are detected as the triggered ones are removed, and only the poses of regarded-clean images are estimated.
In Figure~\ref{figure: defense strip}, we depict the defense performance and surprisingly find that STRIP can only achieve limited defense performance.
Such a high robustness of our IntC attacks benefits from the tiny size of triggers.
Consequently, the triggers are easily covered by perturbations and ignored, and the triggered images will be regarded as clean.

\begin{figure}[!t]
    \centering
    \subfloat[Utility]{
        \centering
        \label{subfigure: defense februus utility}
        \includegraphics[width=0.39\textwidth]{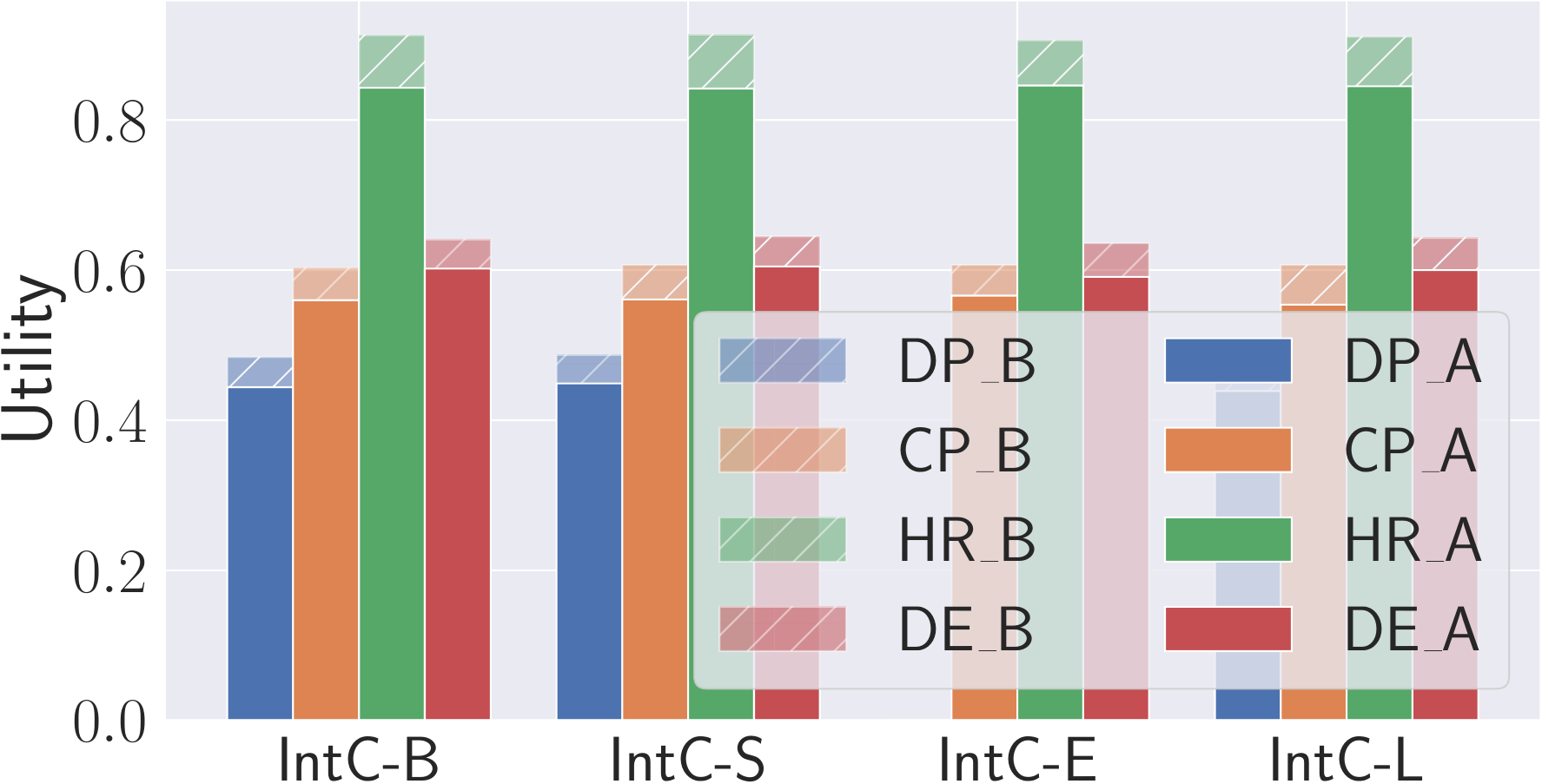}
    }
    \hspace{0.1in}
    \subfloat[ASR]{
        \centering
        \label{subfigure: defense februus asr}
        \includegraphics[width=0.39\textwidth]{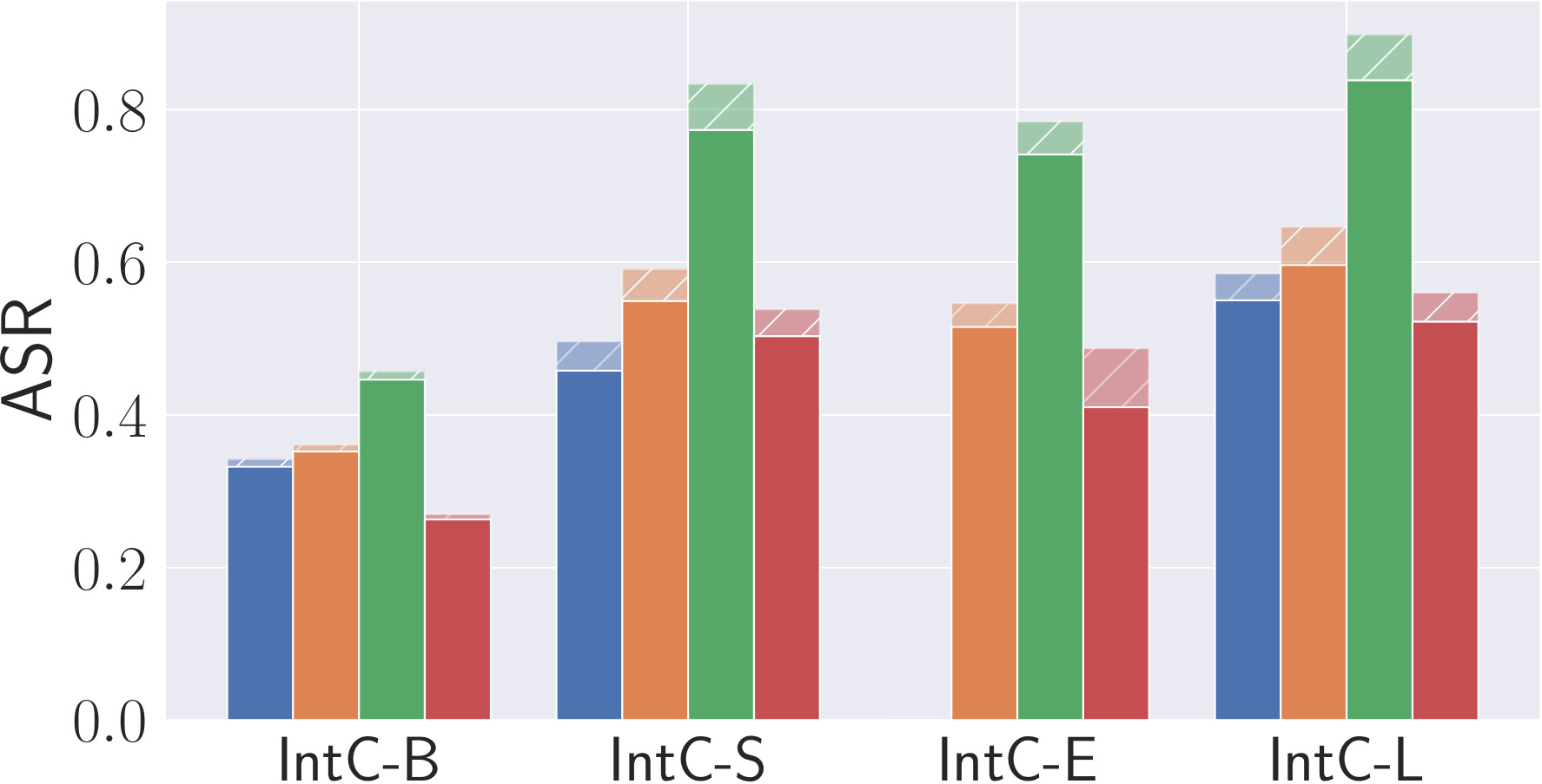}
    }
    \caption{Februus defense performance in terms of Utility and ASR across various HPE tasks.}
    \label{figure: defense februus}
\end{figure}

\shortsection{Februus}
Doan et al. proposed the defense method -- Februus -- to purify query images~\cite{DAR20}.
Specifically, Februus is designed first to inspect and remove triggers and restore the processed images.
To measure the defense performance of Februus, both clean and triggered testing-time images are purified by Februus.
Afterward, the purified testing samples are predicted by the deployed HPE systems.
In Figure~\ref{figure: defense februus}, we present the defense performance of Februus.
Similar to the testing-time purification defense~\cite{ZWZW23} in Section~\ref{subsection: defense}, the ASR and Utility both decrease.
The mitigation of our IntC attacks is as expected by the Februus defense.
On the other hand, the clean images are also modified as the defender has no knowledge of benign information, which will be jeopardized during purification.

\begin{figure}[!t]
    \centering
    \subfloat[Utility]{
        \centering
        \label{subfigure: defense fine-pruning utility}
        \includegraphics[width=0.39\textwidth]{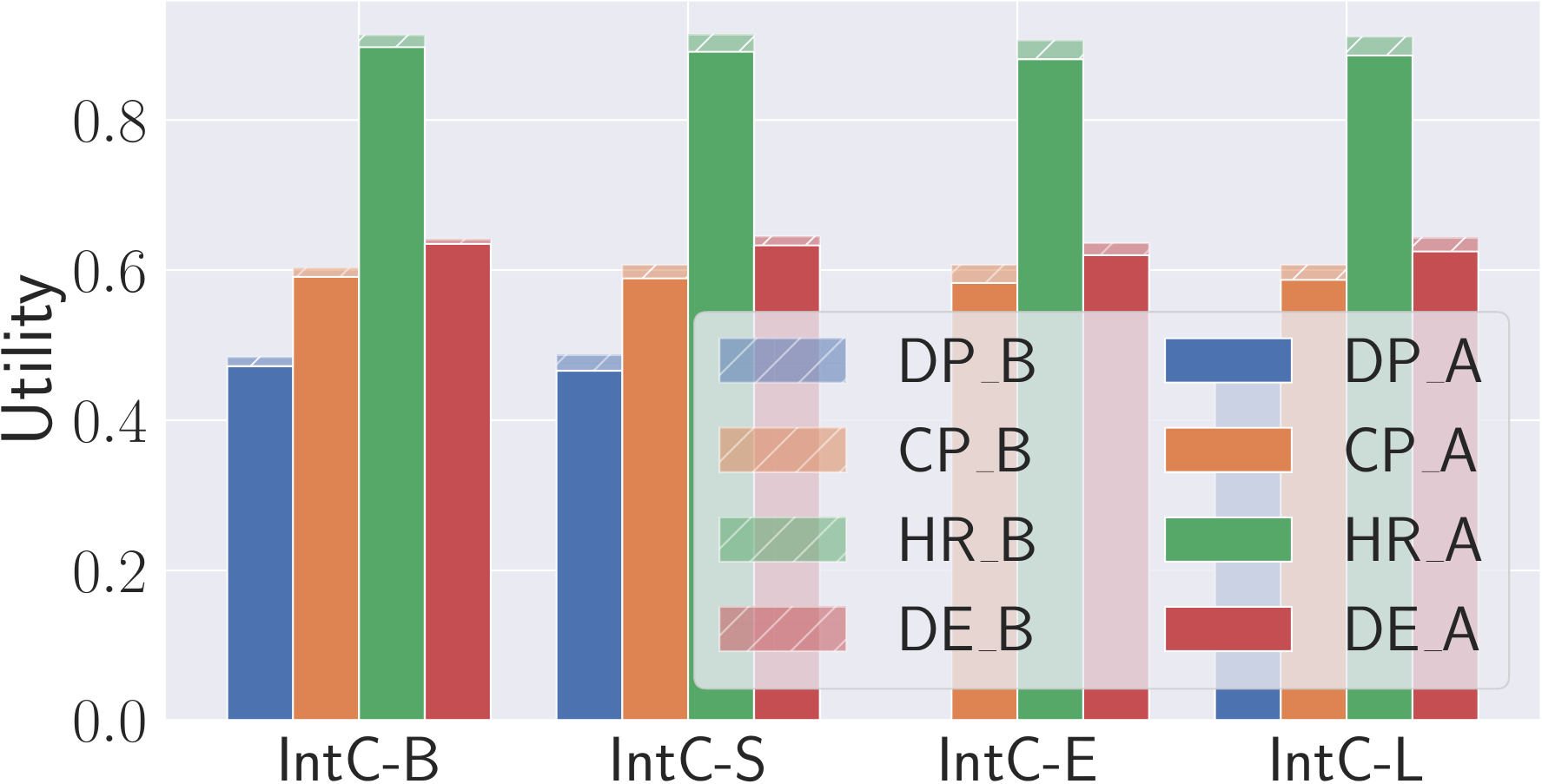}
    }
    \hspace{0.1in}
    \subfloat[ASR]{
        \centering
        \label{subfigure: defense fine-pruning asr}
        \includegraphics[width=0.39\textwidth]{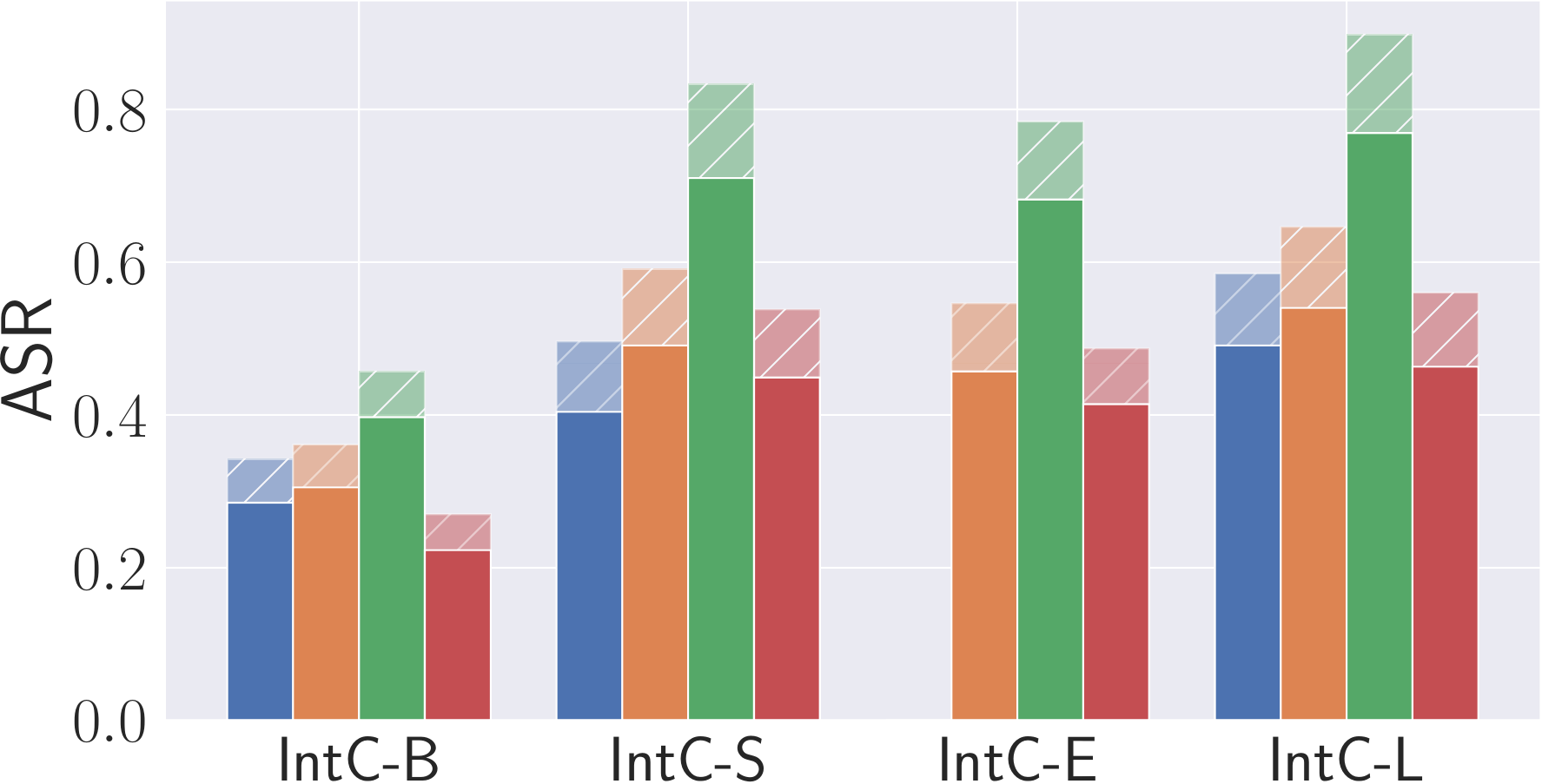}
    }
    \caption{Fine-pruning defense performance in terms of Utility and ASR across various HPE tasks.}
    \label{figure: defense fine-pruning}
\end{figure}

\begin{figure}[!t]
    \centering
    \subfloat[Utility]{
        \centering
        \label{subfigure: defense nad utility}
        \includegraphics[width=0.39\textwidth]{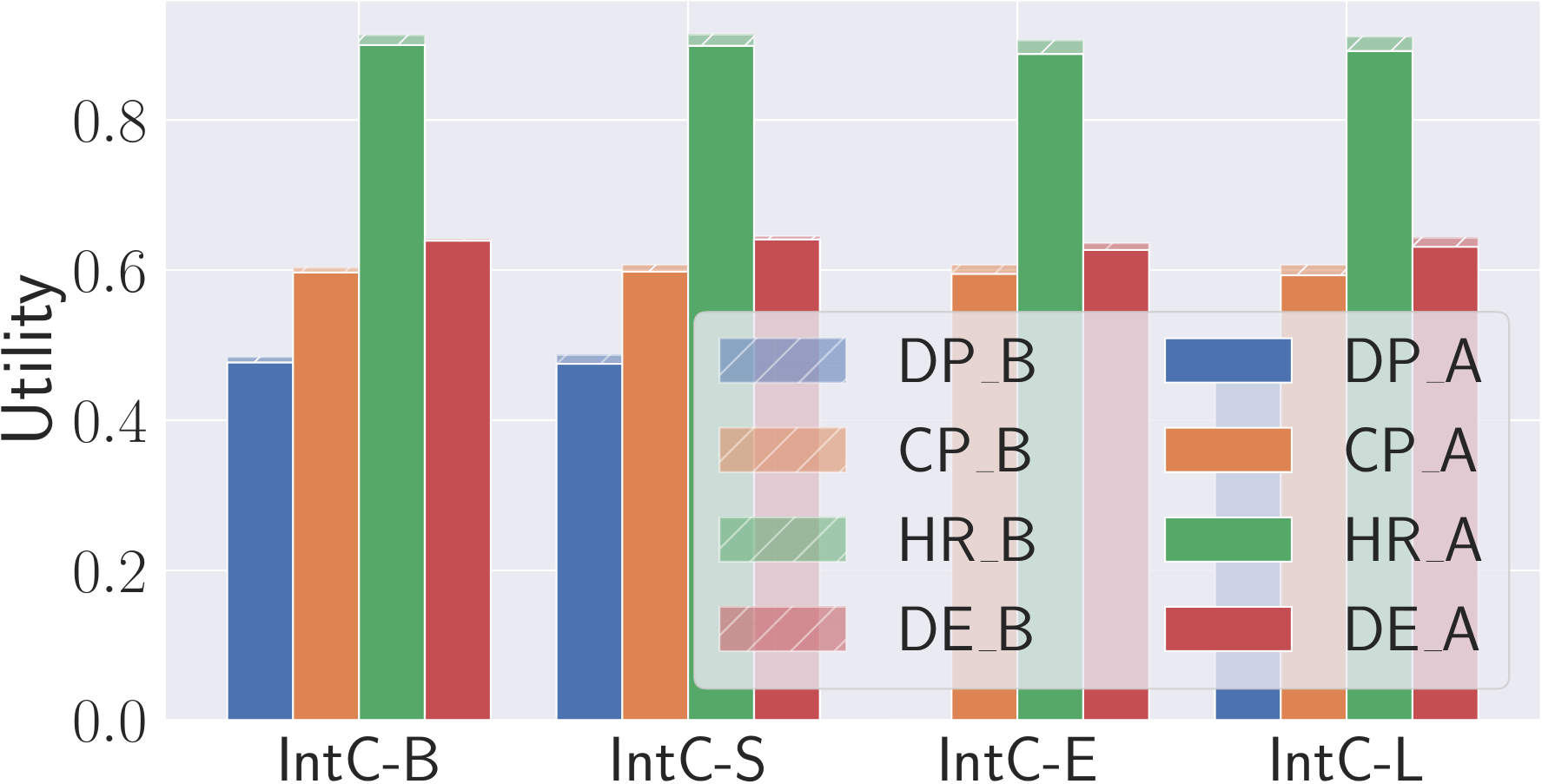}
    }
    \hspace{0.1in}
    \subfloat[ASR]{
        \centering
        \label{subfigure: defense NAD asr}
        \includegraphics[width=0.39\textwidth]{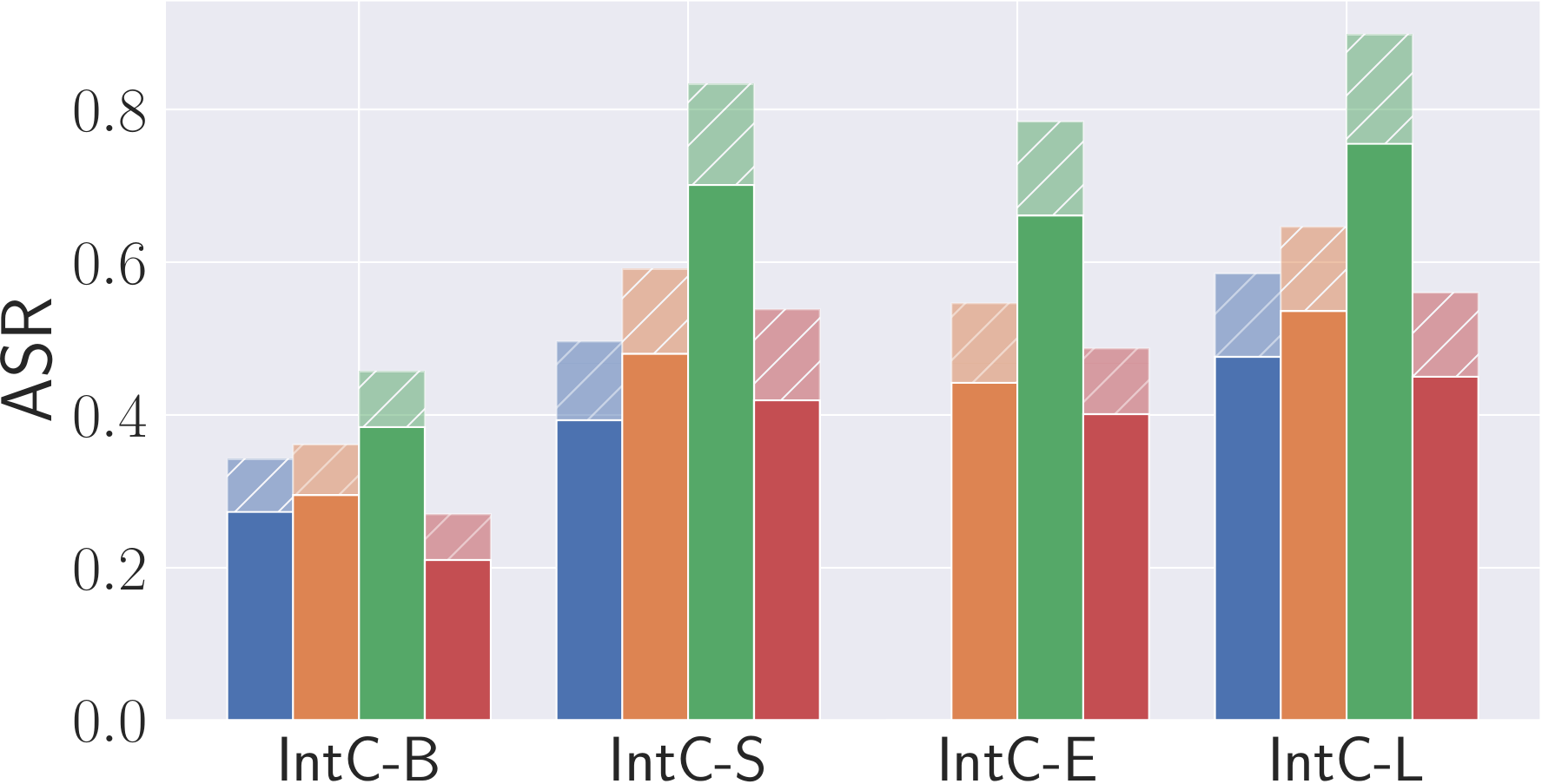}
    }
    \caption{Neural Attention Distillation defense performance in terms of Utility and ASR across various HPE tasks.}
    \label{figure: defense nad}
\end{figure}

\shortsection{Fine-pruning}
Liu et al. proposed the defense method -- Fine-pruning -- to counter backdoor attacks~\cite{LDG18}.
Specifically, Fine-pruning eliminates model neurons that are dormant on clean inputs, aiming to disable backdoor behaviors.
In our evaluation, we first collect a set of clean testing-time images to deploy the fine-pruning defense against IntC attacks.
Then, the HPE model is pruned according to its behavior on this set of clean samples.
Subsequently, the pruned HPE model will provide predictions on other testing-time images of both clean and triggered versions.
We can see from Figure~\ref{figure: defense fine-pruning} that Fine-pruning achieves significant defense performance while maintaining the performance of original HPE tasks.
However, our IntC attacks still outperform the baseline significantly.
Our attack robustness is derived from the fact that the neurons working for clean and triggered images cannot be completely separated.
In other words, even if some model neurons only serve benign information, neurons will always exist for both the clean and triggered ones, and they will not be eliminated.
As our IntC attacks only require a small trigger size and poison number, the remaining neurons can provide good attack performance.
On the other hand, there are two limitations of the Fine-pruning defense:
(1) A set of clean images is needed to eliminate spare model neurons.
This assumption requires a similar effort to guarantee that the training data will not be poisoned;
(2) The pruning process is also time-consuming, especially since HPE normally requires a larger training budget than image classification tasks.
Consequently, addressing the above two concerns might be a potential direction for deriving a better and more practical pruning defense against backdoor attacks in the HPE domain.

\shortsection{Neural Attention Distillation}
Li et al. propose the defense of Neural Attention Distillation (NAD)~\cite{LLKLLM21} to erase backdoor triggers.
Specifically, NAD involves a teacher network to guide the fine-tuning of a backdoored student network on a small subset of clean training data.
To measure the robustness of our IntC attacks against the NAD defense, we conduct the standard fine-tuning on the backdoored HPE model using a small subset of clean training data to gain the teacher HPE model, which follows the steps of the original paper.
We can see from Figure~\ref{figure: defense nad} that NAD achieves nearly perfect protection of Utility and also greatly mitigates our IntC attacks but does not completely eliminate the influence of our attacks.
Such a good defense performance is derived from the attention distillation step.
However, similar to Fine-pruning, there are two limitations hindering the applicable scope of NAD.
To be more concrete, the double training budget is needed to support the establishment of both teacher and student HPE models, which vastly decrease the attacker's ability.
Furthermore, unlike Fine-tuning, which does not assume the source of auxiliary clean images, NAD requires that the clean fine-tuning data be sampled from the training data.
In that case, the clean images in the training data are supposed to be distinguishable, which is the core challenge in against-backdoor defense works that have not been satisfactorily solved.
Consequently, we still believe our IntC attacks can bring security risks in practical scenarios, which encourages future work to propose more effective and realistic defenses.

\end{document}